\documentclass[nojss]{jss}\usepackage[]{graphicx}\usepackage[]{xcolor}
\makeatletter
\def\maxwidth{ %
  \ifdim\Gin@nat@width>\linewidth
    \linewidth
  \else
    \Gin@nat@width
  \fi
}
\makeatother

\usepackage{Sweave}

\usepackage{orcidlink, lmodern}

\newcommand{\class}[1]{`\code{\detokenize{#1}}'}
\newcommand{\fct}[1]{\code{\detokenize{#1}()}}

\usepackage{amsmath, microtype, booktabs}

\usepackage{tikz}
\usetikzlibrary{positioning, arrows.meta, shapes.geometric}
\tikzset{%
  semithick,
  >={Stealth[width=1.5mm,length=2mm]},
  obs/.style n args = {2}{name = #1, circle, draw, inner sep = 8pt, label = center:$#2$}
}

\newcommand{\doo}{\mathrm{do}}
\newcommand*\diff{\mathop{}\!\mathrm{d}}

\author{Santtu Tikka~\orcidlink{0000-0003-4039-4342}\\University of Jyväskyl\"{a}
   \And Jouni Helske~\orcidlink{0000-0001-7130-793X}\\University of Turku}
\Plainauthor{Santtu Tikka, Jouni Helske}

\title{\pkg{dynamite}: An \proglang{R} Package for Dynamic Multivariate Panel Models}
\Plaintitle{dynamite: An R Package for Dynamic Multivariate Panel Models}

\Abstract{
\pkg{dynamite} is an \proglang{R} package for Bayesian inference of intensive panel (time series) data comprising multiple measurements per multiple individuals measured in time. The package supports joint modeling of multiple response variables, time-varying and time-invariant effects, a wide range of discrete and continuous distributions, group-specific random effects, latent factors, and customization of prior distributions of the model parameters. Models in the package are defined via a user-friendly formula interface, and estimation of the posterior distribution of the model parameters takes advantage of state-of-the-art Markov chain Monte Carlo methods. The package enables efficient computation of both individual-level and aggregated predictions and offers a comprehensive suite of tools for visualization and model diagnostics.
}

\Plainkeywords{Bayesian modeling, causal inference, intervention, panel data, prediction, splines}

\Address{
  Santtu Tikka\\
  Department of Mathematics Statistics\\
  Faculty of Mathematics and Science\\
  University of Jyv\"{a}skyl\"{a}\\
  P.O.Box 35, FI-40014, Finland\\
  E-mail: \email{santtu.tikka@jyu.fi}\\
  URL: \url{http://users.jyu.fi/~santikka/}\\

  Jouni Helske\\
  INVEST Research Flagship Centre\\
  University of Turku\\
  FI-20500, Finland\\
  E-mail: \email{jouni.helske@utu.fi}\\
  URL: \url{https://jounihelske.netlify.app/}
}
\IfFileExists{upquote.sty}{\usepackage{upquote}}{}
\begin{document}

\section{Introduction} \label{sec:intro}

Panel data is common in various fields such as social sciences. These data consist of multiple individuals followed over several time points, and there are often many observations per individual at each time, for example, family status and income of each individual at each time point of interest. Such data can be analyzed in various ways, depending on the research questions and the characteristics of the data such as the number of individuals and time points, and the assumed distribution of the response variables. In social sciences, popular, somewhat overlapping modeling approaches include dynamic panel models, fixed effect models, dynamic structural equation models \citep{Asparouhov2018}, cross-lagged panel models (CLPM), and their various extensions such as CLPM with fixed or random effects \citep{arellano1991, Allison2009, Bollen2010, Allison2017, Hamaker2015, Mulder2021} and general cross-lagged panel model \citep{Zyphur2020}.

There are several \proglang{R} \citep{R} packages available from the Comprehensive \proglang{R} Archive Network (CRAN) focusing on the analysis of panel data. The \pkg{plm} package \citep{plm} provides various estimation methods and tests for linear panel data models, while the \pkg{fixest} package \citep{fixest} supports multiple fixed effects and different distributions of response variables. The \pkg{panelr} package \citep{panelr} contains tools for panel data manipulation and estimation methods for so-called ``within-between'' models that combine fixed effect and random effect models. This is done by using \pkg{lme4}, \pkg{geepack}, and \pkg{brms} packages as a backend \citep{lme4, geepack, brms}. The \pkg{lavaan} package \citep{lavaan} provides methods for general structural equation modeling (SEM) and thus can be used to estimate various panel data models such as CLPMs with fixed or random intercepts. Similarly, it is also possible to use general multilevel modeling packages such as \pkg{lme4} and \pkg{brms} directly for panel data modeling. Of these, only \pkg{lavaan} and \pkg{brms} support joint modeling of multiple interdependent response variables, which is typically necessary for multi-step predictions and long-term causal effect estimation \citep{dmpm}.

In traditional panel data models such as the ones supported by the aforementioned packages, the number of time points considered is often assumed to be relatively small, say less than 10, while the number of individuals can be hundreds or thousands \citep{Wooldridge2010}. This is especially true for commonly used ``wide format'' SEM approaches that are unable to consider a large number of time points \citep{Asparouhov2018}. Perhaps due to the small number of time points, the effects of covariates are typically assumed to be time-invariant, although some extensions to time-varying effects have emerged \citep[e.g.,][]{Sun2009, Asparouhov2018, hayakawa2019}. On the other hand, when the number of time points is moderate or large, say hundreds or thousands (sometimes referred to as intensive longitudinal data), it can be reasonable to assume that the dynamics of the system change over time, for example in the form of time-varying effects.

Modeling time-varying effects in (generalized) linear models can be based on state-space models \citep[SSMs,][]{harvey1982, durbin2012, helske2022}, for which there are various \proglang{R} implementations such as \pkg{walker} \citep{helske2022}, \pkg{shrinkTVP} \citep{knaus2021}, and \pkg{CausalImpact} \citep{brodersen2014}. However, these implementations are restricted to a non-panel setting of a single individual and a single response variable. Other approaches include methods based on varying coefficients models \citep{hastie1993, eubank2004}, implemented in \pkg{tvReg} and \pkg{tvem} packages \citep{casas2022, dziak2021}. While \pkg{tvem} supports multiple individuals, it does not support multiple response variables per individual. The \pkg{tvReg} package supports only univariate single-individual responses. Also based on SSMs and differential equations, the \pkg{dynr} package \citep{dynr} provides methods for modeling multivariate dynamic regime-switching models with linear or non-linear latent dynamics and linear-Gaussian observations. Because both multilevel models and SEMs can be defined as SSMs \citep[see e.g.,][]{Sallas1981,KFAS,Chow2010}, other packages supporting general SSMs could be suitable for panel data analysis in principle as well, such as \pkg{KFAS} \citep{KFAS}, \pkg{bssm} \citep{bssm}, and \pkg{pomp} \citep{pomp}. However, SSMs are often computationally demanding especially for non-Gaussian observations where the marginal likelihood is analytically intractable, and a large number of individuals can be problematic, particularly in the presence of additional group-level random effects which complicates the construction of the corresponding state space model \citep{KFAS}.

The \pkg{dynamite} package \citep{dynamitepackage} provides an alternative approach to panel data inference which avoids some of the limitations and drawbacks of the aforementioned methods. First, the dynamic multivariate panel data models (DMPMs), introduced by \citet{dmpm} and implemented in the \pkg{dynamite} package support estimation of effects that vary smoothly over time according to Bayesian P-splines \citep{lang2004}, with penalization based on random walk priors. This allows modeling for example the effects of interventions that increase or decrease over time. Second, \pkg{dynamite} supports a wide variety of distributions for the response variables such as Gaussian, Poisson, binomial, and categorical distributions. Third, with \pkg{dynamite}, we can model an arbitrary number of simultaneous measurements per individual. Finally, the estimation is fully Bayesian using Markov chain Monte Carlo (MCMC) simulation via \proglang{Stan} \citep{Stan} leading to transparent and interpretable quantification of parameter and predictive uncertainty. A comprehensive comparison between DMPMs and other panel data modeling approaches can be found in \citep{dmpm}.

One of the most defining features of \pkg{dynamite} is its high-performance prediction functionality, which is fully automated, supports multi-step predictions over the entire observed time interval, and can operate at the individual level or group level. This is in stark contrast to packages such as \pkg{brms} where, in the presence of lagged response variables as covariates, obtaining such predictions necessitates the computation of manual stepwise predictions and can pose a challenge even for an experienced user. Furthermore, by jointly modeling all endogenous variables simultaneously, \pkg{dynamite} allows us to consider the long-term effects of interventions that take into account the interdependence of the variables in the model.

The paper is organized as follows. In Section~\ref{sec:model} we introduce the dynamic multivariate panel model which is the class of models considered in the \pkg{dynamite} package and describe the assumptions made in the package with respect to these models. Section~\ref{sec:dynamitepackage} introduces the software package and its core features along with two illustrative examples using a real dataset and a synthetic dataset. Sections~\ref{sec:construction} and \ref{sec:fitting} provide a more comprehensive and technical overview of how to define and estimate models using the package. The use of the model fit objects for prediction is discussed in Section~\ref{sec:prediction}. Finally, Section~\ref{sec:summary} summarizes our contributions and provides some concluding remarks.

\section{The dynamic multivariate panel model} \label{sec:model}

Consider an individual \(i\) at time \(t\) with observations \(y_{t,i} = (y_{1,t,i},\ldots, y_{C,t,i})\), \(t=1,\ldots,T\), \(i = 1,\ldots,N\). In other words, at each time point \(t\) we have \(C\) observations from \(N\) individuals, where \(C\) is the number of different response variables that have been measured. The response variables can be univariate or multivariate. We assume that each element of \(y_{t,i}\) can depend on the past observations \(y_{t-\ell,i}\), \(\ell=1,\ldots, t-1\) (where the set of past values can be different for each response) and also on additional exogenous covariates \(x_{t,i}\). In addition, each response variable \(y_{c,t,i}\) can depend on other observations at the same time point \(t\), i.e., the elements of \(y_{t,i}\), with the following restriction. We assume that the response variables can be ordered so that the distribution of \(y_{t,i}\) factorizes according to an ordering \(\pi\) of the responses. We denote the observations at the same time point before observation \(y_{c,t,i}\) in this ordering by \(y_{\pi(c),t,i}\). Thus, the conditional distribution of response \(c\) is completely defined by the observations at the same time point before the response in the ordering \(\pi\), past observations, exogenous covariates, and the model parameters for all \(c = 1,\ldots,C\). For simplicity of the presentation, we now assume that all response variables are univariate and that the responses only depend on the previous time points, i.e., \(\ell = 1\) for all response variables. The set of all model parameters is denoted by \(\theta\). We treat the first \(L\) time points as fixed data, where \(L\) is the highest order of lag dependence in the model. Now, assuming that the elements of \(y_{t,i}\) are conditionally independent given \(y_{t-1,i}\), \(x_{t,i}\), and \(\theta\) we have
\begin{equation} \label{eq:factorization}
  y_{t,i} \sim p_t(y_{t,i} | y_{1:t-1,i},x_{t,i},\theta) = \prod_{c = 1}^C p_{c,t}(y_{c,t,i} | y_{\pi(c),t,i}, y_{1:t-1,i}, x_{t,i}, \theta),
\end{equation}
where \(y_{1:t-1,i}\) denotes the past values of all response variables \((y_{1,i},\ldots,y_{t-1,i})\). Importantly, the parameters of the conditional distributions \(p_{c,t}\) can be time-dependent, enabling us to consider the evolution of the dynamics of our system over time.

Given a suitable link function depending on our distributional assumptions, we define a linear predictor \(\eta_{c,t,i}\) for the conditional distribution \(p_{c,t}\) of each response \(c\) with the following general form:
\begin{equation}\label{eq:linpred}
  \eta_{c,t,i} = \alpha_{c,t} + u^\top_{c,t,i} \beta_c + w^\top_{c,t,i} \delta_{c,t} + z^\top_{c,t,i} \nu_{c,i} + \lambda^\top_{c,i} \psi_{c,t},
\end{equation}
where \(\alpha_{c,t}\) is the (possibly time-varying) common intercept term, \(u^\top_{c,t,i}\) defines the covariates corresponding to the vector of time-invariant coefficients \(\beta_c\), and similarly \(w^\top_{c,t,i}\) defines the covariates for the time-varying coefficients \(\delta_{c,t}\). The term \(z^\top_{c,t,i} \nu_{c,i}\) corresponds to individual-specific random effects, where \(\nu_{1,i},\ldots, \nu_{C,i}\) are assumed to follow a zero-mean Gaussian distribution, either with a diagonal or a full covariance matrix. Note that the covariates in \(u^\top_{c,t,i}\), \(w^\top_{c,t,i}\), and \(z^\top_{c,t,i}\) may contain values of other response variables at the same time point that appear before response \(c\) in the ordering \(\pi\), past observations of the response variables (or transformations of them), or exogenous covariates. Covariates in \(z^\top_{c,t,i}\) can overlap those in \(u^\top_{c,t,i}\) and \(w^\top_{c,t,i}\) resulting in an interpretation for \(\nu_{c,i}\) that corresponds to individual-specific deviations from the population-level effects \(\beta_c\) and \(\delta_{c,t}\), respectively. In contrast, the covariates in \(u^\top_{c,t,i}\) and \(w^\top_{c,t,i}\) should in general not overlap to ensure the identifiability of their respective model parameters. The final term \(\lambda^\top_{c,i} \psi_{c,t}\) is a product of latent individual loadings \(\lambda_{c,i}\) and a univariate latent dynamic factor \(\psi_{c,t}\). The latent factors can be correlated between responses.

For the time-varying coefficients \(\delta_{c,t}\) (and similarly for time-varying \(\alpha_{c,t}\) and the latent factor \(\psi_{c,t}\)), we use Bayesian P-splines \citep[penalized B-splines,][]{eilers1996, lang2004} such that
\[
  \delta_{c,t,k} = b^\top_t \omega_{c,k}, \quad k=1,\ldots,K,
\]
where \(K\) is the number of covariates, \(b_t\) is a vector of B-spline basis function values at time \(t\), and \(\omega_{c,k}\) is a vector of corresponding spline coefficients. We assume a B-spline basis of equally spaced knots on the time interval from \(L+1\) to \(T\) with \(D\) degrees of freedom. In general, the number of B-splines \(D\) used for constructing the splines for the study period \(1,\ldots,T\) can be chosen freely, but the actual value is not too important \citep[as long as \(D\) is larger than the degree of the spline, e.g., three for cubic splines,][]{Wood2020}. Therefore, we use the same \(D\) basis functions for all time-varying effects. To mitigate overfitting due to too large a value of \(D\), we define a random walk prior \citep{lang2004} for \(\omega_{c,k}\) as
\[
  \omega_{c,k,1} \sim p(\omega_{c,k,1}), \quad
  \omega_{c,k,d} \sim N(\omega_{c,k,d-1}, \tau^2_{c,k}), \quad d=2, \ldots, D,
\]
with a user-defined prior \(p(\omega_{c,k,1})\) on the first coefficient, which due to the structure of \(b_1\) corresponds to a prior on \(\delta_{c,k,1}\). Here, the parameter \(\tau_{c,k}\) controls the smoothness of the spline curves. While the different time-varying coefficients are modeled as independent a~priori, the latent factors \(\psi_{c,t}\) can be modeled as correlated via correlated spline coefficients \(\omega_{c,k}\). See Appendix~\ref{app:latent_factor} for the details of the parametrization of the latent factor term.

For categorical, multivariate, and other distributions with multiple dimensions or components, we can extend the definition of the linear predictor in Equation~\ref{eq:linpred} to account for each dimension by simply replacing the index \(c\) with indices \(c,s\) where \(s\) denotes the index of the dimension, \(s = 1,\ldots,S(c)\), and \(S(c)\) is the number of dimensions of response \(c\). This extension also applies to the spline coefficients.

\section[The dynamite package]{The \pkg{dynamite} package}\label{sec:dynamitepackage}

The \pkg{dynamite} package provides an easy-to-use interface for fitting DMPMs in \proglang{R}. As the package is part of rOpenSci (\url{htps://ropensci.org}), it complies with its rigorous software standards and the development version of \pkg{dynamite} can be installed from the \proglang{R}-universe system \url{https://ropensci.org/r-universe/}. The stable version of the package is available from CRAN at \url{https://cran.r-project.org/package=dynamite}. The software is published under the GNU general public license (GPL \(\geq\) 3) and can be obtained in \proglang{R} by running the following commands:
\begin{CodeChunk}
\begin{CodeInput}
R> install.packages("dynamite")
R> library("dynamite")
\end{CodeInput}
\end{CodeChunk}
The package takes advantage of several other well-established \proglang{R} packages. Estimation of the models is carried out by \proglang{Stan} for which both \pkg{rstan} and \pkg{cmdstanr} interfaces are available \citep{rstan, cmdstanr}. More specifically, the MCMC simulation uses the No-U-Turn sampler \citep[NUTS,][]{hoffman2014} which is an extension of Hamiltonian Monte Carlo \citep[HMC,][]{Neal2011}. The \pkg{data.table} package \citep{datatable} is used for efficient computation and memory management of predictions and internal data manipulations. For posterior inference and visualization, \pkg{ggplot2} and \pkg{posterior} packages are leveraged \citep{ggplot2, posterior}. Leave-one-out (LOO) and leave-future-out (LFO) cross-validation methods are implemented with the help of the \pkg{loo} package \citep{loo}. All of the aforementioned dependencies are available on CRAN except for \pkg{cmdstanr} whose installation is optional and needed only if the user wishes to use the \pkg{CmdStan} backend for \proglang{Stan}. Although not required for \pkg{dynamite}, we also install the \pkg{dplyr}, \pkg{pder}, and \pkg{pryr} packages \citep{dplyr, pder, pryr}, as we will use them in the subsequent sections. In addition to the required \proglang{R} packages, \pkg{dynamite} also requires \proglang{C++} compilation capabilities due to \proglang{Stan}. Specifically for Windows users, this means that RTools has to be installed (\url{https://cran.r-project.org/bin/windows/Rtools/}).

Several example datasets and corresponding model fit objects are included in \pkg{dynamite} which are used throughout this paper for illustrative purposes. The script files to generate these datasets and the model fit objects can be found in the package GitHub repository (\url{https://https://github.com/ropensci/dynamite/}) under the \code{data-raw} directory. Table~\ref{tab:dynamite_funs} provides an overview of the available functions and methods of the package. Before presenting the technical details, we demonstrate the key features of the package and the general workflow by performing an illustrative analysis on a real dataset and a synthetic dataset.

\begin{table}[!ht]
  \footnotesize
  \centering
  \begin{tabular}{lll}
    \toprule
    Function & Output & Description \\
    \midrule
    \multicolumn{3}{l}{\emph{Model fitting}} \\[0.15cm]
    \fct{dynamite} & \class{dynamitefit} & Estimate a dynamic multivariate panel model \\
    \fct{dynamice} & \class{dynamitefit} & Estimate a DMPM with multiple imputation \\
    \midrule
    \multicolumn{3}{l}{\emph{Model formula construction}} \\[0.15cm]
    \fct{dynamiteformula}   & \class{dynamiteformula} & Define a response variable \\
    \fct{+.dynamiteformula} & \class{dynamiteformula} & Add definitions to a model formula \\
    \fct{obs}               & \class{dynamiteformula} & Define a response variable (alias) \\
    \fct{aux}               & \class{dynamiteformula} & Define a deterministic variable \\
    \fct{splines}           & \class{splines}         & Define P-splines for time-varying coefficients \\
    \fct{random_spec}       & \class{random_spec}     & Define additional properties of random effects \\
    \fct{lags}              & \class{lags}            & Define lagged covariates for all responses \\
    \fct{lfactor}           & \class{lfactor}         & Define latent factors \\
    \midrule
    \multicolumn{3}{l}{\proglang{S3} \emph{Methods for} \class{dynamitefit} \emph{objects}} \\[0.15cm]
    \fct{as.data.frame}       & \class{tbl_df}      & Extract posterior samples or summaries \\
    \fct{as.data.table}       & \class{data.table}  & Extract posterior samples or summaries \\
    \fct{as_draws}            & \class{draws_df}    & Extract posterior samples or summaries \\
    \fct{as_draws_df}         & \class{draws_df}    & Extract posterior samples or summaries \\
    \fct{coef}                & \class{tbl_df}      & Extract posterior samples or summaries \\
    \fct{confint}             & \class{matrix}      & Extract credible intervals \\
    \fct{fitted}              & \class{data.table}  & Compute fitted values \\
    \fct{formula}             & \class{language}    & Extract the model formula \\
    \fct{get_code}            & \class{data.frame}  & Extract the \proglang{Stan} model code* \\
    \fct{get_data}            & \class{list}        & Extract the data used to fit the model* \\
    \fct{get_parameter_dims}  & \class{list}        & Extract parameter dimensions* \\
    \fct{get_parameter_names} & \class{character}   & Extract parameter names \\
    \fct{get_parameter_types} & \class{character}   & Extract parameter types \\
    \fct{get_priors}          & \class{data.frame}  & Extract the prior distribution definitions* \\
    \fct{hmc_diagnostics}     & \class{dynamitefit} & Compute HMC diagnostics \\
    \fct{lfo}                 & \class{lfo}         & Compute LFO cross-validation for the model \\
    \fct{loo}                 & \class{loo}         & Compute LOO cross-validation for the model \\
    \fct{mcmc_diagnostics}    & \class{dynamitefit} & Compute MCMC diagnostics \\
    \fct{ndraws}              & \class{integer}     & Extract the number of posterior draws \\
    \fct{nobs}                & \class{integer}     & Extract the number of observations \\
    \fct{plot}                & \class{ggplot}      & Visualize posterior distributions \\
    \fct{predict}             & \class{data.frame}  & Compute predictions \\
    \fct{print}               & \class{dynamitefit} & Print information on the model fit* \\
    \fct{summary}             & \class{data.frame}  & Print a summary of the model fit \\
    \fct{update}              & \class{dynamitefit} & Update the model fit \\
    \bottomrule
  \end{tabular}
  \caption{The functionality of \pkg{dynamite}. Asterisks denote \class{dynamitefit} methods that are also available for \class{dynamiteformula} objects.}
  \label{tab:dynamite_funs}
\end{table}

\subsection{Bayesian inference of seat belt usage and traffic fatalities}\label{sec:seatbelt}

As the first illustration, we consider the effect of seat belt laws on traffic fatalities using data from the \pkg{pder} package, originally analyzed by \citet{Cohen2003}. The data consists of the number of traffic fatalities and other related variables in the United States from all 51 states for every year from 1983 to 1997. During this time, many states passed laws regarding mandatory seat belt use. We distinguish two types of laws: secondary enforcement law and primary enforcement law. Secondary enforcement means that the police can fine violators only when they are stopped for other offenses, whereas in primary enforcement the police can also stop and fine based on the seat belt use violation itself. This dataset is named \code{SeatBelt} and it can be loaded into the current \proglang{R} session by running:
\begin{CodeChunk}
\begin{CodeInput}
R> data("SeatBelt", package = "pder")
\end{CodeInput}
\end{CodeChunk}
To begin, we rename some variables and compute additional transformations to make the subsequent analyses straightforward.
\begin{CodeChunk}
\begin{CodeInput}
R> library("dplyr")
R> seatbelt <- SeatBelt |>
+    mutate(
+      miles = (vmturban + vmtrural) / 10000,
+      log_miles = log(miles),
+      fatalities = farsocc,
+      income10000 = percapin / 10000,
+      law = factor(
+        case_when(
+          dp == 1 ~ "primary",
+          dsp == 1 ~ "primary",
+          ds == 1 & dsp == 0 ~ "secondary",
+          TRUE ~ "no_law"
+        ),
+        levels = c("no_law", "secondary", "primary")
+      )
+    )
\end{CodeInput}
\end{CodeChunk}
We are interested in the effect of the seat belt law on traffic fatalities in terms of car occupants via the changes in seat belt usage. For this purpose, we build a joint model for seat belt usage and fatalities. We model the rate of seat belt usage with a beta distribution (with a logit link) and assume that the usage depends on the level of the seat belt law, state-level effects (modeled as random intercepts), and overall time-varying trend (modeled as a spline), which captures potential changes in the general tendency to use a seat belt in the US. We model the number of fatalities with a negative binomial distribution (with a log link) using the total miles traveled as an offset. In addition to the seat belt usage and state-level random intercepts, we also use several other variables related to traffic density, speed limit, alcohol usage, and income (see \code{?pder::SeatBelt} for details) as controls. First, we construct the model formula that defines the distributions of the response variables, their covariates, and the splines used for the time-varying effects:
\begin{CodeChunk}
\begin{CodeInput}
R> seatbelt_formula <-
+    obs(usage ~ -1 + law + random(~1) + varying(~1), family = "beta") +
+    obs(fatalities ~ usage + densurb + densrur +
+      bac08 + mlda21 + lim65 + lim70p + income10000 + unemp + fueltax +
+      random(~1) + offset(log_miles), family = "negbin") +
+    splines(df = 10)
\end{CodeInput}
\end{CodeChunk}
In the code above, we used \code{random(~1)} to define group-specific random effects, \code{varying(~1)} to define a time-varying intercept term, and \code{splines(df = 10)} to define the degrees of freedom for the splines of the time-varying intercept. These components and other functionality of \pkg{dynamite} related to defining models are described at length in Section~\ref{sec:construction}. Next, we fit the model
\begin{CodeChunk}
\begin{CodeInput}
R> fit <- dynamite(
+    dformula = seatbelt_formula,
+    data = seatbelt, time = "year", group = "state",
+    chains = 4, cores = 4, seed = 0, refresh = 0
+  )
\end{CodeInput}
\end{CodeChunk}
We note that fitting the model takes several minutes, which is common when using MCMC methods. Compiling the model also contributes to the total time taken, and sampling from precompiled models is generally faster. Sampling time can be reduced by leveraging parallelization, as we have done here by setting \code{chains = 4} and \code{cores = 4}. Parallel capabilities of \pkg{dynamite} are discussed at greater length in Section~\ref{sec:fitting}.

We can extract the estimated coefficients with the \fct{summary} method which shows clear positive effects for both secondary enforcement and primary enforcement laws:
\begin{CodeChunk}
\begin{CodeInput}
R> summary(fit, types = "beta", response = "usage") |>
+    select(parameter, mean, sd, q5, q95)
\end{CodeInput}
\begin{CodeOutput}
# A tibble: 2 x 5
  parameter                mean     sd    q5   q95
  <chr>                   <dbl>  <dbl> <dbl> <dbl>
1 beta_usage_lawsecondary 0.495 0.0463 0.417 0.571
2 beta_usage_lawprimary   1.05  0.0820 0.919 1.19 
\end{CodeOutput}
\end{CodeChunk}
While these coefficients can be interpreted as changes in log-odds as usual, we also estimate the marginal means using the \fct{fitted} method which returns the posterior samples of the expected values of the responses at each time point given the covariates. For this purpose, we create a new data frame for each level of the \code{law} factor and assign every state to uphold this particular law. We then call \fct{fitted} using these data, compute the averages of over the states and finally over the posterior samples:
\begin{CodeChunk}
\begin{CodeInput}
R> seatbelt_new <- seatbelt
R> seatbelt_new$law[] <- "no_law"
R> pnl <- fitted(fit, newdata = seatbelt_new)
R> seatbelt_new$law[] <- "secondary"
R> psl <- fitted(fit, newdata = seatbelt_new)
R> seatbelt_new$law[] <- "primary"
R> ppl <- fitted(fit, newdata = seatbelt_new)
R> bind_rows(no_law = pnl, secondary = psl, primary = ppl, .id = "law") |>
+    mutate(
+      law = factor(law, levels = c("no_law", "secondary", "primary"))
+    ) |>
+    group_by(law, .draw) |>
+    summarize(mm = mean(usage_fitted)) |>
+    group_by(law) |>
+    summarize(
+      mean = mean(mm),
+      q5 = quantile(mm, 0.05),
+      q95 = quantile(mm, 0.95)
+    )
\end{CodeInput}
\begin{CodeOutput}
# A tibble: 3 x 4
  law        mean    q5   q95
  <fct>     <dbl> <dbl> <dbl>
1 no_law    0.359 0.347 0.372
2 secondary 0.468 0.458 0.477
3 primary   0.591 0.567 0.615
\end{CodeOutput}
\end{CodeChunk}
These estimates are in line with the results of \citet{Cohen2003} who report the law effects on seat belt usage as increases of 11 and 22 percentage points for secondary enforcement and primary enforcement laws, respectively.

For the effect of seat belt laws on the number of traffic fatalities, we compare the number of fatalities with 68\% seat belt usage against 90\% usage. These values, coinciding with the national average in 1996 and the target of 2005, were also used by \citet{Cohen2003} who reported an increase in annual lives saved as 1500--3000. We do this by comparing the differences in total fatalities across states for each year, and by averaging over the years, again with the help of the \fct{fitted} method;
\begin{CodeChunk}
\begin{CodeInput}
R> seatbelt_new <- seatbelt
R> seatbelt_new$usage[] <- 0.68
R> p68 <- fitted(fit, newdata = seatbelt_new)
R> seatbelt_new$usage[] <- 0.90
R> p90 <- fitted(fit, newdata = seatbelt_new)
R> bind_rows(low = p68, high = p90, .id = "usage") |>
+    group_by(year, .draw) |>
+    summarize(
+      s = sum(
+        fatalities_fitted[usage == "low"] -
+          fatalities_fitted[usage == "high"]
+      )
+    ) |>
+    group_by(.draw) |>
+    summarize(m = mean(s)) |>
+    summarize(
+      mean = mean(m),
+      q5 = quantile(m, 0.05),
+      q95 = quantile(m, 0.95)
+    )
\end{CodeInput}
\begin{CodeOutput}
# A tibble: 1 x 3
   mean    q5   q95
  <dbl> <dbl> <dbl>
1 1561.  773. 2322.
\end{CodeOutput}
\end{CodeChunk}
In this example, the model did not contain any lagged responses as covariates, so it was enough to compute predictions for each time point essentially independently using the \fct{fitted} method. However, when the responses depend on the past values of themselves or of other responses, as is the case for example in cross-lagged panel models, estimating long-term causal effects such as \(\E(y_{t+k} | \doo(y_t)), k = 1,\ldots\), where \(\doo(y_t)\) denotes an intervention on \(y_t\) \citep{Pearl2009}, is more complicated. We illustrate this in our next example.

\subsection{Causal effects in a multivariate model}\label{sec:multichannel}

We consider the following simulated multivariate data available in the \pkg{dynamite} package and the estimation of causal effects.
\begin{CodeChunk}
\begin{CodeInput}
R> head(multichannel_example)
\end{CodeInput}
\begin{CodeOutput}
  id time          g  p b
1  1    1 -0.6264538  5 1
2  1    2 -0.2660091 12 0
3  1    3  0.4634939  9 1
4  1    4  1.0451444 15 1
5  1    5  1.7131026 10 1
6  1    6  2.1382398  8 1
\end{CodeOutput}
\end{CodeChunk}
The data contains 50 unique groups (variable \code{id}), over 20 time points (\code{time}), a continuous variable \(g_t\) (\code{g}), a variable with non-negative integer
values \(p_t\) (\code{p}), and a binary variable \(b_t\) (\code{b}). We define the following model (which actually matches the data-generating process used to generate the data):
\begin{CodeChunk}
\begin{CodeInput}
R> multi_formula <- obs(g ~ lag(g) + lag(logp), family = "gaussian") +
+    obs(p ~ lag(g) + lag(logp) + lag(b), family = "poisson") +
+    obs(b ~ lag(b) * lag(logp) + lag(b) * lag(g), family = "bernoulli") +
+    aux(numeric(logp) ~ log(p + 1) | init(0))
\end{CodeInput}
\end{CodeChunk}
Here, the \fct{aux} function creates a deterministic transformation of \(p_t\) defined as \(\log(p_t + 1)\) which can subsequently be used for other responses as a covariate and correctly computes the transformation for predictions. Because the model also contains a lagged value of \code{logp}, we define the initial value of \code{logp} to be 0 at the first time point via the \fct{past} declaration. Without the initial value, we would receive a warning message when fitting the model, but in this case we could safely ignore the warning because the model contains lags of \code{b} and \code{g} as well meaning that the first time point in the model is treated as fixed and does not enter the model fitting process. This makes the \fct{past} declaration redundant in this instance, but it is good practice to always define the initial values of deterministic variables when the model contains their lagged values to avoid accidental \code{NA} values when the variable is evaluated. A directed acyclic graph (DAG) that depicts the causal relationships of the variables in the model is shown in Figure~\ref{fig:multichannel_dag}. We fit the model using the \fct{dynamite} function.
\begin{figure}[!ht]
  \begin{center}
    \begin{tikzpicture}[scale=0.6]
    \node [obs = {g_{t-1}}{g_{t-1}}] at (-6, 3) {\(\vphantom{0}\)};
    \node [obs = {p_{t-1}}{p_{t-1}}] at (-6, 0) {\(\vphantom{0}\)};
    \node [obs = {b_{t-1}}{b_{t-1}}] at (-6,-3) {\(\vphantom{0}\)};
    \node [obs = {g_{t}}{g_{t}}] at (-2, 3) {\(\vphantom{0}\)};
    \node [obs = {p_{t}}{p_{t}}] at (-2, 0) {\(\vphantom{0}\)};
    \node [obs = {b_{t}}{b_{t}}] at (-2,-3) {\(\vphantom{0}\)};
    \node [obs = {g_{t+1}}{g_{t+1}}] at (2, 3) {\(\vphantom{0}\)};
    \node [obs = {p_{t+1}}{p_{t+1}}] at (2, 0) {\(\vphantom{0}\)};
    \node [obs = {b_{t+1}}{b_{t+1}}] at (2,-3) {\(\vphantom{0}\)};
    \path [->] (g_{t-1}) edge (g_{t});
    \path [->] (g_{t-1}) edge (p_{t});
    \path [->] (g_{t-1}) edge (b_{t});
    \path [->] (p_{t-1}) edge (g_{t});
    \path [->] (p_{t-1}) edge (p_{t});
    \path [->] (p_{t-1}) edge (b_{t});
    \path [->] (b_{t-1}) edge (p_{t});
    \path [->] (b_{t-1}) edge (b_{t});
    \path [->] (g_{t}) edge (g_{t+1});
    \path [->] (g_{t}) edge (p_{t+1});
    \path [->] (g_{t}) edge (b_{t+1});
    \path [->] (p_{t}) edge (g_{t+1});
    \path [->] (p_{t}) edge (p_{t+1});
    \path [->] (p_{t}) edge (b_{t+1});
    \path [->] (b_{t}) edge (p_{t+1});
    \path [->] (b_{t}) edge (b_{t+1});
    \end{tikzpicture}
  \end{center}
  \caption{A directed acyclic graph for the multivariate model with arrows corresponding to the assumed direct causal effects. A cross-section at times \(t-1\), \(t\), and \(t+1\) is shown. The vertices and edges corresponding to the deterministic transformation \(\log(p_t + 1)\) are projected out for clarity.}
  \label{fig:multichannel_dag}
\end{figure}
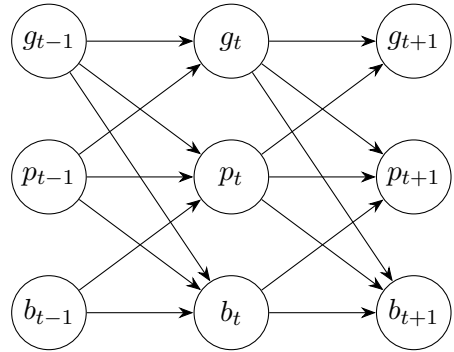
\begin{CodeChunk}
\begin{CodeInput}
R> multichannel_fit <- dynamite(
+    dformula = multi_formula,
+    data = multichannel_example, time = "time", group = "id",
+    chains = 4, cores = 4, seed = 0, refresh = 0
+  )
\end{CodeInput}
\end{CodeChunk}
We can obtain posterior samples or summary statistics of the model using the \fct{as.data.frame}, \fct{coef}, and \fct{summary} methods, but here we opt for visualizing the results as depicted in Figure~\ref{fig:multichannel_betas} by using the \fct{plot} method:
\begin{CodeChunk}
\begin{CodeInput}
R> library("ggplot2")
R> theme_set(theme_bw())
R> plot(multichannel_fit, types = "beta") +
+    labs(title = "")
\end{CodeInput}
\begin{figure}
\includegraphics[width=\maxwidth]{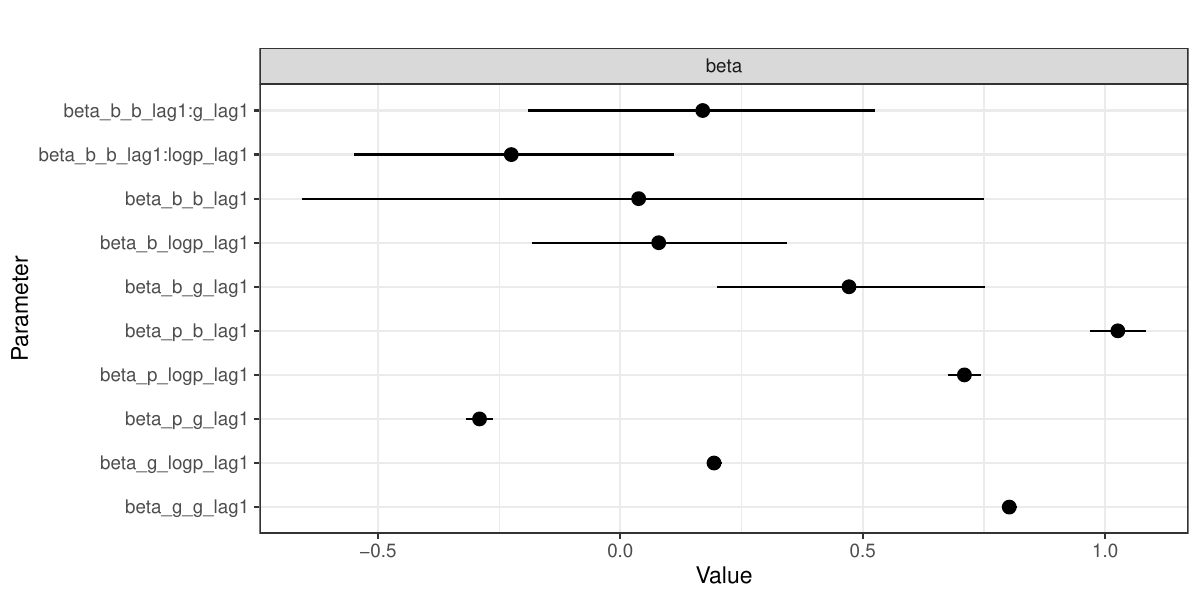} \caption[Posterior means and 90\% posterior intervals of the time-invariant coefficients for the multivariate model]{Posterior means and 90\% posterior intervals of the time-invariant coefficients for the multivariate model.}\label{fig:multichannel_betas}
\end{figure}
\end{CodeChunk}
Note the naming of the model parameters; for example, \code{beta_b_g_lag1} corresponds to a time-invariant coefficient \code{beta} for response \code{b} of the lagged covariate \code{g}.

Assume now that we are interested in the causal effect of \(b_5\) on \(g_t\) at times \(t = 6, \ldots, 20\). There is no direct effect from \(b_5\) to \(g_6\), but because \(g_t\) affects \(b_{t+1}\) (and \(p_{t+1}\)), which in turn affects all variables at \(t+2\), we should see an indirect effect of \(b_5\) to \(g_t\) from time \(t = 7\) onward. For this task, we first create a new dataset where the values of our response variables after time \(t = 5\) are assigned to be missing.
\begin{CodeChunk}
\begin{CodeInput}
R> multichannel_newdata <- multichannel_example |>
+    mutate(across(g:b, ~ ifelse(time > 5, NA, .x)))
\end{CodeInput}
\end{CodeChunk}
We then obtain predictions for time points \(t = 6,\ldots,20\) when \(b_t\) is assigned to be 0 or 1 for every individual at time \(t = 5\), corresponding to the interventions \(\doo(b_5 = 0)\) and \(\doo(b_5 = 1)\).
\begin{CodeChunk}
\begin{CodeInput}
R> new0 <- multichannel_newdata |>
+    mutate(b = ifelse(time == 5, 0, b))
R> pred0 <- predict(multichannel_fit, newdata = new0, type = "mean")
R> new1 <- multichannel_newdata |>
+    mutate(b = ifelse(time == 5, 1, b))
R> pred1 <- predict(multichannel_fit, newdata = new1, type = "mean")
\end{CodeInput}
\end{CodeChunk}
By default, the output from \fct{predict} is a single data frame containing the original new data and the samples from the posterior predictive distribution of new observations. By defining \code{type = "mean"}, we specify that we are interested in the posterior distribution of the expected values instead. In this case, the predicted values in the output are in the columns \code{g_mean}, \code{p_mean}, and \code{b_mean} where the \code{NA} values of the \code{newdata} argument are replaced with the posterior predictive samples from the model (the output also contains an additional column corresponding to the auxiliary response \code{logp} and posterior draw index variable \code{.draw}).
\begin{CodeChunk}
\begin{CodeInput}
R> head(pred0, n = 10) |>
+    round(3)
\end{CodeInput}
\begin{CodeOutput}
   id time .draw g_mean p_mean b_mean  logp      g  p  b
1   1    1     1     NA     NA     NA 1.792 -0.626  5  1
2   1    2     1     NA     NA     NA 2.565 -0.266 12  0
3   1    3     1     NA     NA     NA 2.303  0.463  9  1
4   1    4     1     NA     NA     NA 2.773  1.045 15  1
5   1    5     1     NA     NA     NA 2.398  1.713 10  0
6   1    6     1  1.858  3.716  0.723 1.946     NA NA NA
7   1    7     1  1.944  6.750  0.760 1.946     NA NA NA
8   1    8     1  1.787  2.696  0.720 0.693     NA NA NA
9   1    9     1  1.723  2.797  0.780 1.609     NA NA NA
10  1   10     1  1.660  5.784  0.730 2.079     NA NA NA
\end{CodeOutput}
\end{CodeChunk}
We can now compute summary statistics over the individuals and then over the posterior samples to obtain the posterior distribution of the expected causal effects \(\E(g_t | \doo(b_5))\) as
\begin{CodeChunk}
\begin{CodeInput}
R> sumr <- list(b0 = pred0, b1 = pred1) |>
+    bind_rows(.id = "case") |>
+    group_by(case, .draw, time) |>
+    summarize(mean_t = mean(g_mean)) |>
+    group_by(case, time) |>
+    summarize(
+      mean = mean(mean_t),
+      q5 = quantile(mean_t, 0.05, na.rm = TRUE),
+      q95 = quantile(mean_t, 0.95, na.rm = TRUE)
+    )
\end{CodeInput}
\end{CodeChunk}
It is also possible to perform the marginalization over groups within \fct{predict} by using the \code{funs} argument, which can be used to provide a named list of lists of functions to be applied for the corresponding response. This approach can save a considerable amount of memory in case of a large number of observations and groups. The names of the outermost list should be names of response variables. The output is now returned as a \class{list} with two components, \code{simulated} and \code{observed}, with the new samples and the original \code{newdata} respectively. In our case, we can write
\begin{CodeChunk}
\begin{CodeInput}
R> pred0b <- predict(
+    multichannel_fit, newdata = new0, type = "mean",
+    funs = list(g = list(mean_t = mean))
+  )$simulated
R> pred1b <- predict(
+    multichannel_fit, newdata = new1, type = "mean",
+    funs = list(g = list(mean_t = mean))
+  )$simulated
R> sumrb <- list(b0 = pred0b, b1 = pred1b) |>
+    bind_rows(.id = "case") |>
+    group_by(case, time) |>
+    summarize(
+      mean = mean(mean_t_g),
+      q5 = quantile(mean_t_g, 0.05, na.rm = TRUE),
+      q95 = quantile(mean_t_g, 0.95, na.rm = TRUE)
+    )
\end{CodeInput}
\end{CodeChunk}
The resulting data frame \code{sumrb} is equal to the previous \code{sumr} (apart from stochasticity due to the simulation of new trajectories). We can then visualize our predictions as shown in Figure~\ref{fig:multichannelvisual} by writing:
\begin{CodeChunk}
\begin{CodeInput}
R> ggplot(sumr, aes(time, mean)) +
+    geom_ribbon(aes(ymin = q5, ymax = q95), alpha = 0.5, na.rm = TRUE) +
+    geom_line(na.rm = TRUE) +
+    scale_x_continuous(n.breaks = 10) +
+    facet_wrap(~ case)
\end{CodeInput}
\begin{figure}
\includegraphics[width=\maxwidth]{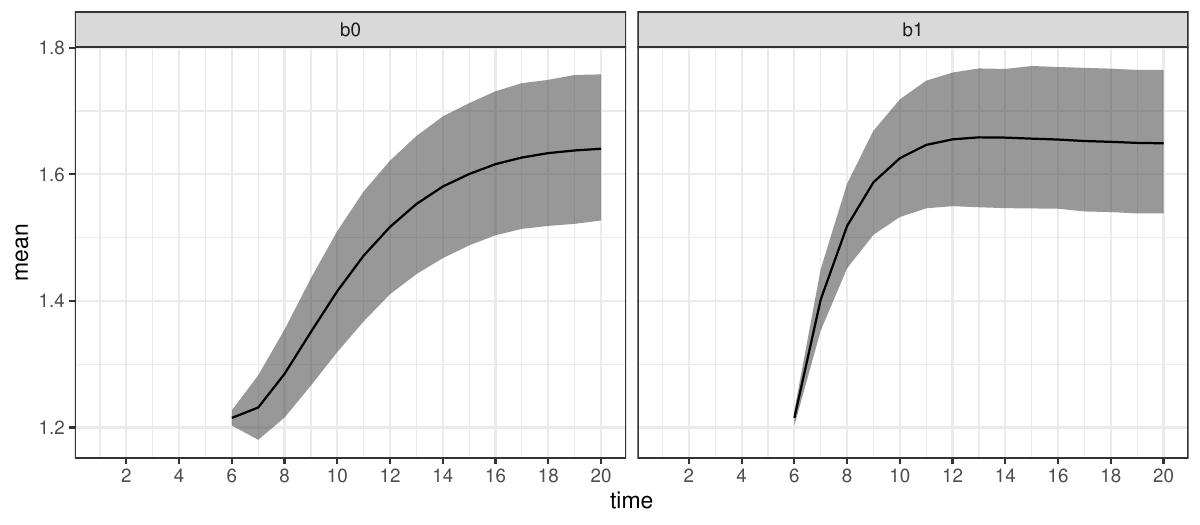} \caption[Expected causal effects of interventions \(\doo(b_5 = 0)\) and \(\doo(b_5 = 1)\) on \(g_t\)]{Expected causal effects of interventions \(\doo(b_5 = 0)\) and \(\doo(b_5 = 1)\) on \(g_t\). The black lines show the posterior means and the gray areas show 90\% posterior intervals.}\label{fig:multichannelvisual}
\end{figure}
\end{CodeChunk}
Predictions for the first 5 time points in \code{sumr} are \code{NA} for all groups by design because our new data supplied to the \fct{predict} method for both interventions contained observations for those time points, which is why we set \code{na.rm = TRUE} to avoid a warning in the above code. Note that these estimates do indeed coincide with the causal effects (assuming of course that our model is correct), as we can apply the backdoor adjustment formula \citep{Pearl1995} to obtain the expected causal effect:
\[
  \E(g_t | \doo(b_5 = x)) = \int \E(g_t | b_5 = x, g_5, p_5)\Prob(g_5, p_5)\diff g_5 \diff p_5,
\]
where the integral over \(p_5\) should be understood as a sum as \(p_5\) is discrete. In the code above, \code{mean_t} is the estimate of this expected value. In addition, we compute an estimate of the difference
\[
  \E(g_t | \doo(b_5 = 1)) - \E(g_t | \doo(b_5 = 0)),
\]
to directly compare the effects of the interventions by writing:
\begin{CodeChunk}
\begin{CodeInput}
R> sumr_diff <- list(b0 = pred0, b1 = pred1) |>
+    bind_rows(.id = "case") |>
+    group_by(.draw, time) |>
+    summarize(
+      mean_t = mean(g_mean[case == "b1"] - g_mean[case == "b0"])
+    ) |>
+    group_by(time) |>
+    summarize(
+      mean = mean(mean_t),
+      q5 = quantile(mean_t, 0.05, na.rm = TRUE),
+      q95 = quantile(mean_t, 0.95, na.rm = TRUE)
+    )
\end{CodeInput}
\end{CodeChunk}
We can also plot the difference between the expected causal effects as shown in Figure~\ref{fig:multichannelcausaldiffplot} by running:
\begin{CodeChunk}
\begin{CodeInput}
R> ggplot(sumr_diff, aes(time, mean)) +
+    geom_ribbon(aes(ymin = q5, ymax = q95), alpha = 0.5, na.rm = TRUE) +
+    geom_line(na.rm = TRUE) +
+    scale_x_continuous(n.breaks = 10)
\end{CodeInput}
\begin{figure}
\includegraphics[width=\maxwidth]{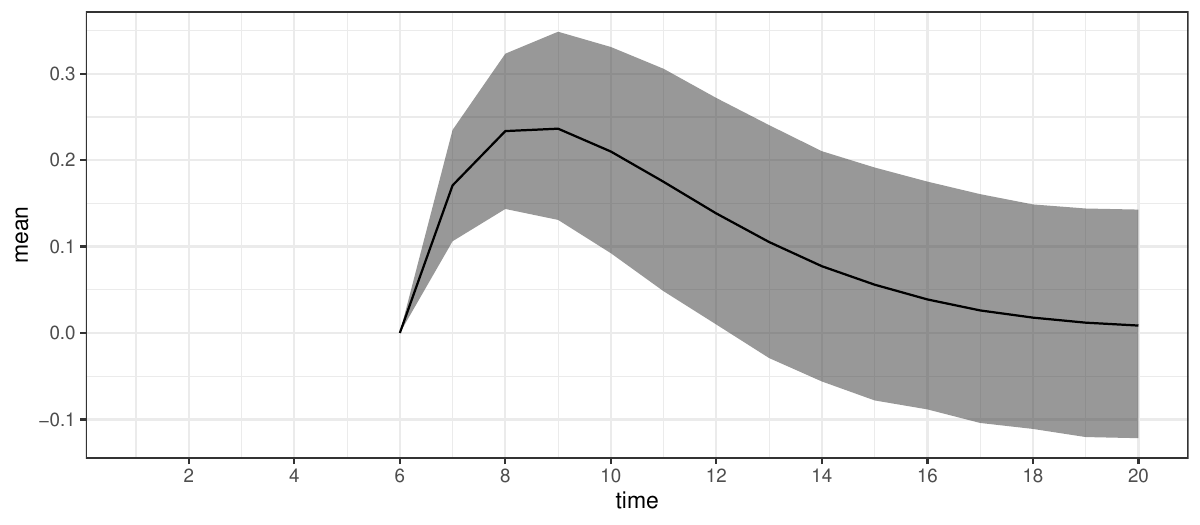} \caption[Difference between the expected causal effects \(\E(g_t | \doo(b_5 = 1)) - \E(g_t | \doo(b_5 = 0)) \)]{Difference between the expected causal effects \(\E(g_t | \doo(b_5 = 1)) - \E(g_t | \doo(b_5 = 0)) \). The black line shows the posterior mean and the gray area shows a 90\% posterior interval.}\label{fig:multichannelcausaldiffplot}
\end{figure}
\end{CodeChunk}
This shows that there is a short-term effect of \(b_5\) on \(g_t\) where the size of the effect diminishes towards zero in time, although the posterior uncertainty is quite large.

\section{Model construction}\label{sec:construction}

Here we describe the various model components that can be included in the model formulas of the \pkg{dynamite} package. These components are modular and easily combined in any order via a specialized \code{+} operator while ensuring that the model formula is well-defined and syntactically valid before estimating the model. The model formula components define the response variables, auxiliary response variables, the splines used for time-varying coefficients, correlated random effects, and latent factors.

\subsection{Defining response variables}\label{sec:defining}

The response variables are defined by combining the response-specific formulas defined via the function \fct{dynamiteformula} for which a shorthand alias \fct{obs} is also provided. We will henceforth use this alias for brevity. The function \fct{obs} takes three arguments: \code{formula}, \code{family}, and \code{link} which define how the response variable depends on the covariates in the standard \proglang{R} formula syntax, the family of the response variable as a \class{character} string, and the link function to use as a \class{character} string, respectively. The link function specification is optional with each \code{family} having a default link. The response-specific definitions are combined into a single model definition with the \code{+} operator of \class{dynamiteformula} objects. For example, the following formula
\begin{CodeChunk}
\begin{CodeInput}
R> dform <- obs(y ~ lag(x), family = "gaussian") +
+    obs(x ~ z, family = "poisson")
\end{CodeInput}
\end{CodeChunk}
defines a model with two responses. First, we declare that \code{y} is a Gaussian response variable depending on the previous value of \code{x} (\code{lag(x)}). Next, we add a second response declaring \code{x} as Poisson distributed depending on an exogenous variable \code{z} (for which we do not define any distribution). Recalling the seat belt usage example from Section~\ref{sec:seatbelt}, we wrote
\begin{Code}
obs(usage ~ -1 + law + random(~1) + varying(~1), family = "beta") +
obs(fatalities ~ usage + densurb + densrur +
  bac08 + mlda21 + lim65 + lim70p + income10000 + unemp + fueltax +
  random(~1) + offset(log_miles), family = "negbin")
\end{Code}
which defines the seat belt usage (\code{usage}) as Beta-distributed and the traffic fatalities (\code{fatalities}) as negative binomial distributed. Note that the model formula can be defined without referencing any external data, just like an \proglang{R} formula can. The model formula is an object of class \class{dynamiteformula} for which the \fct{print} method provides a summary of the defined response variables, including the response variable names, families and formulas, and other model components:
\begin{CodeChunk}
\begin{CodeInput}
R> print(dform)
\end{CodeInput}
\begin{CodeOutput}
  Family   Formula   
y gaussian y ~ lag(x)
x poisson  x ~ z     
\end{CodeOutput}
\end{CodeChunk}
Currently, the package supports the following distributions for the observations:

\begin{description}
\item[Bernoulli] (\code{"bernoulli"}) with logit link.
\item[Beta] (\code{"beta"}) with logit link, using mean and precision parametrization.
\item[Binomial] (\code{"binomial"}) with logit link.
\item[Categorical] (\code{"categorical"}) with a softmax link using the first category as the reference. It is recommended to use \proglang{Stan} version~2.23 or higher which enables the use of the  \code{categorical_logit_glm} function in the generated \proglang{Stan} code for improved computational performance. See the documentation of \code{categorical_logit_glm} in the \proglang{Stan} function reference manual \url{https://mc-stan.org/users/documentation/} for further information.
\item[Exponential] (\code{"exponential"}) with log link.
\item[Gamma] (\code{"gamma"}) with log link, using mean and shape parametrization.
\item[Gaussian] (\code{"gaussian"}) with identity link, parameterized using mean and standard deviation.
\item[Multinomial] (\code{"multinomial"}) with a softmax link using the first category as the reference.
\item[Multivariate Gaussian] (\code{"mvgaussian"}) with identity link for each dimension, parameterized using the mean vector, the standard deviation vector, and the Cholesky decomposition of the correlation matrix.
\item[Negative binomial] (\code{"negbin"}) with log link, using mean and dispersion parametrization, with an optional known offset variable. See the documentation of the \fct{NegBinomial2} function in the \proglang{Stan} function reference manual.
\item[Ordered] (\code{"cumulative"}) with logit or probit link for ordinal regression using cumulative parametrization for the class probabilities.
\item[Poisson] (\code{"poisson"}) with log link, with an optional known offset variable.
\item[Student \(t\)] (\code{"student"}) with identity link, parameterized using location, scale, and degrees of freedom.
\end{description}

There is also a special response variable type \code{"deterministic"} which can be used to define deterministic transformations of other variables in the model. This special type is explained in greater detail in Section~\ref{sec:auxiliary}.

\subsection{Lagged responses and covariates}\label{sec:lags}

Models in the \pkg{dynamite} package have limited support for contemporaneous dependencies to avoid complex cyclic dependencies that would render the processing of missing data, subsequent predictions, and causal inference challenging or impossible. In other words, the model structure must be acyclic in a sense that there is an order of the response variables such that each response at time \(t\) can be unambiguously defined in this order in terms of responses that have already been defined at time \(t\) or in terms of other variables in the model at time \(t-1\) as formulated in Equation~\ref{eq:factorization}. The acyclicity of the model implied by the model formula defined by the user is checked automatically upon construction. To demonstrate, the following formula is valid:
\begin{Code}
obs(y ~ x, family = "gaussian") +
  obs(x ~ z, family = "poisson")
\end{Code}
However, if we were to add another model component \code{obs(z ~ y, family = "gaussian")}, then the formula would no longer be valid as \code{y} is defined in terms of \code{x}, \code{x} is defined in terms of \code{z}, and \code{z} is defined in terms of \code{y}, creating a cycle from \code{y} to \code{y}. This type of model formulation would produce an error due to the cyclic definition of the responses. On the other hand, there are no limitations concerning the dependence of response variables and their previous values or previous values of exogenous covariates, i.e., lags. In the first example of Section~\ref{sec:defining}, we used the syntax \code{lag(x)}, a shorthand for \code{lag(x, k = 1)}, which defines a first-order lag of the variable \code{x} to be used as a covariate. Higher-order lags can also be defined by adjusting the argument \code{k}. The argument \code{x} of \fct{lag} can either be a response variable or an exogenous covariate.

The model component \fct{lags} can also be used to quickly add lagged responses as covariates across multiple responses. This component adds a lagged value of each response in the model as a covariate for every response. For example, calling
\begin{Code}
obs(y ~ z, family = "gaussian") +
  obs(x ~ z, family = "poisson") +
  lags(k = 1)
\end{Code}
would add \code{lag(y, k = 1)} and \code{lag(x, k = 1)} as covariates of \code{x} and \code{y}. Therefore, the previous code would produce the same model as writing
\begin{Code}
obs(y ~ z + lag(y, k = 1) + lag(x, k = 1), family = "gaussian") +
  obs(x ~ z + lag(y, k = 1) + lag(x, k = 1), family = "poisson")
\end{Code}
The function \fct{lags} can help to simplify the individual model formulas, especially when the model consists of many responses each having a large number of lags. Just as with the function \fct{lag}, the argument \code{k} in \fct{lags} can be adjusted to add higher-order lags of each response for each response, but for \fct{lags} it can also be a vector so that multiple lags can be added at once. The inclusion of lagged response variables in the model implies that some time points must be considered fixed in the estimation. The number of fixed time points in the model is equal to the highest order lag \(k\) of any observed response variable in the model (defined either via \fct{lag} terms or the model component \fct{lags}). Lags of exogenous covariates do not affect the number of fixed time points, as such covariates are not modeled.

\subsection{Time-varying and time-invariant effects}

The \code{formula} argument of \fct{obs} can also contain a special term \fct{varying}, which defines the time-varying part of the model equation. For example, we could write
\begin{Code}
obs(x ~ z + varying(~ -1 + w), family = "poisson")
\end{Code}
to define a model equation with a time-invariant intercept, a time-invariant effect of \code{z}, and a time-varying effect of \code{w}. We also avoid defining a duplicate intercept by writing \code{-1} within \fct{varying} in order to avoid identifiability issues in the model estimation. Alternatively, we could define a time-varying intercept, in which case we would write:
\begin{Code}
obs(x ~ -1 + z + varying(~ w), family = "poisson")
\end{Code}
The part of the formula not wrapped with \fct{varying} is assumed to correspond to the time-invariant part of the model, which can alternatively be defined with the special syntax \fct{fixed}. This means that the following lines would all produce the same model:
\begin{Code}
obs(x ~ z + varying(~ -1 + w), family = "poisson")
obs(x ~ -1 + fixed(~ z) + varying(~ -1 + w), family = "poisson")
obs(x ~ fixed(~ z) + varying(~ -1 + w), family = "poisson")
\end{Code}
The use of \fct{fixed} is therefore optional in the formula. If both time-varying and time-invariant intercepts are defined, the model will default to using a time-varying intercept and an appropriate warning is provided for the user:
\begin{CodeChunk}
\begin{CodeInput}
R> obs(y ~ 1 + varying(~1), family = "gaussian")
\end{CodeInput}
\begin{CodeOutput}
Warning: Both time-constant and time-varying intercept specified:
i Defaulting to time-varying intercept.
\end{CodeOutput}
\end{CodeChunk}
When defining time-varying effects, we also need to define how their respective regression coefficients depend on time. For this purpose, a \fct{splines} component should be added to the model formula, as we did in the seat belt usage example, where the term \code{splines(df = 10)} defines a cubic B-spline with 10 degrees of freedom for the time-varying coefficients, which corresponds to the time-varying intercept in this instance. If the model contains multiple time-varying coefficients, the same spline basis is used for all coefficients, with unique spline coefficients and their corresponding random-walk standard deviations for each coefficient. The \fct{splines} component constructs the matrix of cardinal B-splines \(B_t\) using the \fct{bs} function of the \pkg{splines} package based on the degrees of freedom (\code{df}) and the degree of the polynomials used to construct the splines (\code{degree}, the default being 3 corresponding to cubic B-splines). It is also possible to switch between centered (the default) and non-centered parametrization \citep{Papaspiliopoulos2007} for the spline coefficients using the \code{noncentered} argument of the \fct{splines} component. This can affect the sampling efficiency of \proglang{Stan}, depending on the model and the informativeness of the data \citep{Betancourt2013}.

\subsection{Group-level random effects}

Random effect terms of a response variable for each group can be defined using the special term \fct{random} within the \code{formula} argument of \fct{obs}, analogously to \fct{varying} and \fct{fixed}. By default, all random effects within a group and across all responses are modeled as zero-mean multivariate Gaussian. The optional model component \fct{random_spec} can be used to define non-correlated random effects as \code{random_spec(correlated = FALSE)}. In addition, as with the spline coefficients, it is possible to switch between centered and non-centered (the default) parametrization of the random effects using the \code{noncentered} argument of \fct{random_spec}.

For example, the following code defines a Gaussian response variable \code{x} with a time-invariant common effect of \code{z} as well as a group-specific intercept and group-specific effect of \code{z}.
\begin{Code}
obs(x ~ z + random(~1 + z), family = "gaussian")
\end{Code}
The variable that defines the groups in the data is provided in the call to the model fitting function \fct{dynamite} via the \code{group} argument as shown in Section~\ref{sec:fitting}. Recalling again the seat belt usage example, we wrote
\begin{Code}
obs(usage ~ -1 + law + random(~1) + varying(~1), family = "beta")
\end{Code}
which defines a group-specific intercept term for the usage, which in this case corresponds to state-level intercepts.

\subsection{Latent factors}

Instead of common time-varying intercept terms, it is possible to define response-specific univariate latent factors using the \fct{lfactor} model component. Each latent factor is modeled as a spline, with degrees of freedom and spline degree defined via the \fct{splines} component (in the case that the model also contains time-varying effects, the same spline basis definition is currently used for both latent factors and time-varying effects). The argument \code{responses} of \fct{lfactor} defines which responses should have a latent factor, while argument \code{correlated} determines whether the latent factors should be modeled as correlated. Again, users can switch between centered and non-centered parametrizations using the argument \code{noncentered_psi}.

In general, dynamic latent factors are not identifiable without imposing some constraints on the factor loadings \(\lambda\) or the latent factor \(\psi\) \citep[see, e.g.,][]{bai2015}, especially in the context of DMPMs and \pkg{dynamite}. In \pkg{dynamite}, these identifiability problems are addressed via internal reparametrization and an additional argument \code{nonzero_lambda} which determines whether we assume that the expected value of the factor loadings is zero or not. The theory and thorough experiments regarding the robustness of these identifiability constraints is a work in progress, so some caution should be used regarding the use of the \fct{lfactor} component.

\subsection{Multivariate responses}\label{sec:multivariate}

While models with more than one response variable are multivariate by definition, it is also possible to define responses that follow multivariate distributions. In \fct{obs}, a multivariate response should be given by specifying the data variables that define its dimensions and combining them with \fct{c}. For instance, suppose that we wish to define a multivariate Gaussian response whose dimensions are given by variables \code{y1}, \code{y2}, and \code{y3} with a time-invariant effect of \code{x} for each dimension. Then we would write:
\begin{Code}
obs(c(y1, y2, y3) ~ x, family = "mvgaussian")
\end{Code}
It is also possible to define a distinct formula for each dimension by separating the dimension-specific definitions with a vertical bar \code{|}, for example
\begin{Code}
obs(c(y1, y2, y3) ~ 1 | x | lag(y1), family = "mvgaussian")
\end{Code}
would define no covariates for the first dimension, \code{x} as a covariates for the second dimension, and the lagged value of the first dimension as a covariate for the third dimension. The dimension-specific formulas can contain time-invariant and time-varying effects, group-specific random effects, and latent factors, just like univariate response formulas can.

\subsection{Number of trials and offset variables}

The special terms \fct{trials} and \fct{offset} define the number of trials for binomial and multinomial responses, and an offset variable for negative binomial and Poisson responses, respectively. The arguments to these special terms can be exogenous covariates or other response variables of the model, as long as the possible contemporaneous dependencies do not violate the acyclicity of the model as described in Section~\ref{sec:lags}. For example, the size of a population could be used as an offset when modeling the prevalence of a disease. Modeling the population size in addition to the prevalence enables future predictions for the prevalence when the future population size is unknown.

Both \fct{trials} and \fct{offset} terms are added to the formula similar to \fct{varying} or \fct{random} terms:
\begin{Code}
obs(y ~ z + trials(n), family = "binomial") +
  obs(x ~ z + offset(w), family = "poisson")
\end{Code}
The code above would define a model with a binomial response \code{y} with a time-invariant effect of \code{z} and the number of trials given by the variable \code{n}, and a Poisson response \code{x} with a time-invariant effect of \code{z} and the variable \code{w} as the offset. In the seat belt model, we used log-miles as an offset for the fatalities as follows
\begin{Code}
obs(fatalities ~ usage + densurb + densrur +
  bac08 + mlda21 + lim65 + lim70p + income10000 + unemp + fueltax +
  random(~1) + offset(log_miles), family = "negbin")
\end{Code}

\subsection{Auxiliary response variables}\label{sec:auxiliary}

In addition to declaring response variables via \fct{obs}, we can also use the function \fct{aux} to define auxiliary responses which are deterministic transformations of other variables in the model. Defining these auxiliary variables explicitly instead of defining them implicitly on the right-hand side of the formulas, i.e., by using the ``as is'' function \fct{I}, makes the subsequent prediction steps clearer and allows easier checks of the model validity. Because of this, we do not allow the use of \fct{I} in the \code{formula} argument of \fct{dynamiteformula}. The values of auxiliary variables are computed automatically when fitting the model, and dynamically during prediction, making the use of lagged values and other transformations possible and automatic in prediction as well. An example of a model formula using an auxiliary response could be
\begin{Code}
obs(y ~ lag(log1x), family = "gaussian") +
  obs(x ~ z, family = "poisson") +
  aux(numeric(log1x) ~ log(1 + x) | init(0))
\end{Code}
For auxiliary responses, the formula declaration via \code{~} should be understood as mathematical equality or assignment, where the right-hand side provides the defining expression of the variable on the left-hand side. Thus, the example above defines an auxiliary response \code{log1x} as the logarithm of \code{1 + x}, and assigns it to be of type \class{numeric}. The type declaration is required, because it might not be possible to unambiguously determine the type of the response variable based on its expression alone from the data, especially if the expression contains \class{factor} type variables. Supported types include \class{factor}, \class{numeric}, \class{integer}, and \class{logical}. A warning is issued to the user if the type declaration is missing from the auxiliary variable definition, and the variable will default to the \class{numeric} type:
\begin{CodeChunk}
\begin{CodeInput}
R> aux(log1x ~ log(1 + x) | init(0))
\end{CodeInput}
\begin{CodeOutput}
Warning: No type specified for deterministic channel `log1x`:
i Assuming type is <numeric>.
\end{CodeOutput}
\end{CodeChunk}
Auxiliary variables can be used directly in the formulas of other responses, just like any other variable. The function \fct{aux} does not use the \code{family} argument, as the \code{family} is automatically set to \code{"deterministic"} which is a special family type of the \fct{obs} function. Note that lagged values of deterministic auxiliary variables do not imply fixed time points. Instead, they must be given starting values using one of the two special syntax variants, \fct{init} or \fct{past} after the main formula separated by the \code{|} symbol.

In the example above, because the formula for \code{y} contains a lagged value of \code{log1x} as a covariate, we also need to supply \code{log1x} with a single initial value that determines the value of the lag at the first time point. Here, \code{init(0)} defines the initial value of \code{lag(log1x)} to be zero for all individuals. In general, if the model contains higher-order lags of an auxiliary variable, then \fct{init} can be supplied with a vector initializing each lag.

While \fct{init} defines the same starting value to be used for all individuals, an alternative, special syntax \fct{past} can be used, which takes an \proglang{R} expression as its argument and computes the starting value for each individual based on that expression. The expression is evaluated in the context of the \code{data} supplied to the model fitting function \fct{dynamite}. For example, instead of \code{init(0)} in the example above, we could write:
\begin{Code}
obs(y ~ lag(log1x), family = "gaussian") +
  obs(x ~ z, family = "poisson") +
  aux(numeric(log1x) ~ log(1 + x) | past(log(z)))
\end{Code}
which defines that the value of \code{lag(log1x)} at the first time point is \code{log(z)} for each individual, using the value of \code{z} in the data to be supplied to compute the actual value of the expression. The special syntax \fct{past} can also be used if the model contains higher-order lags of auxiliary responses. In this case, additional observations from the variables bound by the expression given as the argument will simply be used to define the initial values.

\subsection{Visualizing the model structure}\label{sec:modelvis}

A \fct{plot} method is available for \class{dynamiteformula} objects that can be used to easily visualize the overall model structure as a DAG. This method can produce either a \class{ggplot} object of the model plot or a \class{character} string describing a \pkg{TikZ} \citep{tikz} code to render the figure in a report, for example. As an illustration, we produce an analogous \class{ggplot} version of the DAG depicting the multivariate model that was considered in Section~\ref{sec:multichannel}. Figure~\ref{fig:multichanneldagplot} shows the plots obtained by running the following.
\begin{CodeChunk}
\begin{CodeInput}
R> plot(multi_formula)
R> plot(multi_formula, show_auxiliary = FALSE)
\end{CodeInput}
\begin{figure}
\includegraphics[width=0.5\linewidth]{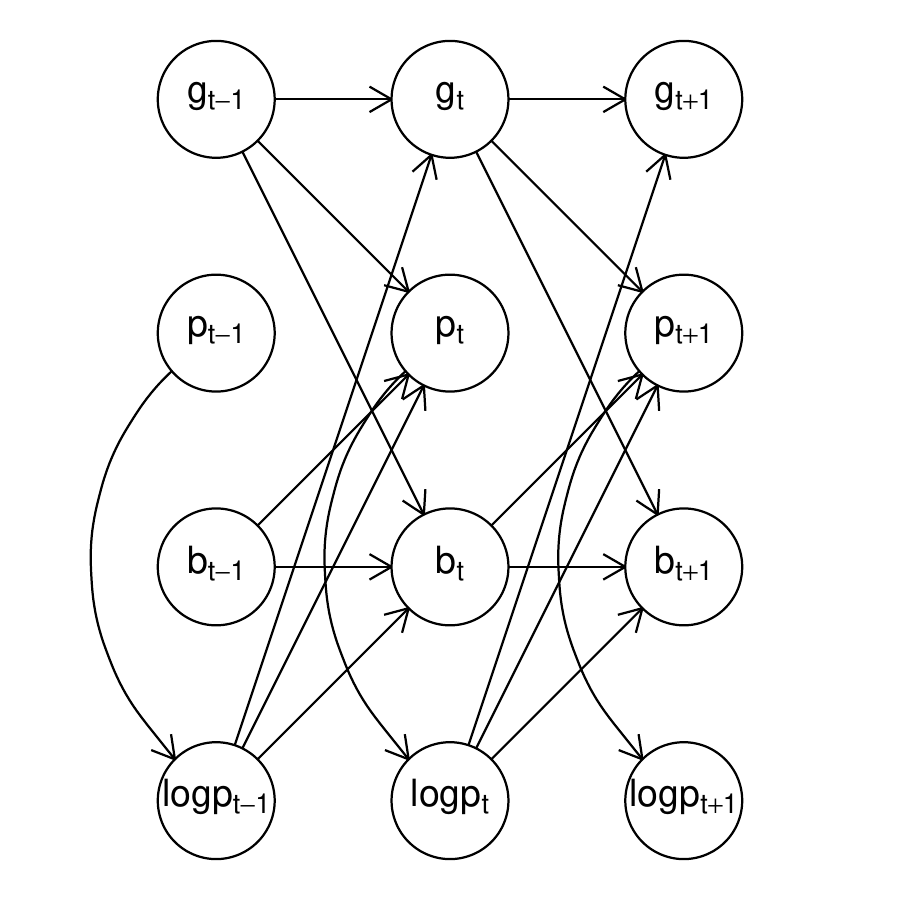} \includegraphics[width=0.5\linewidth]{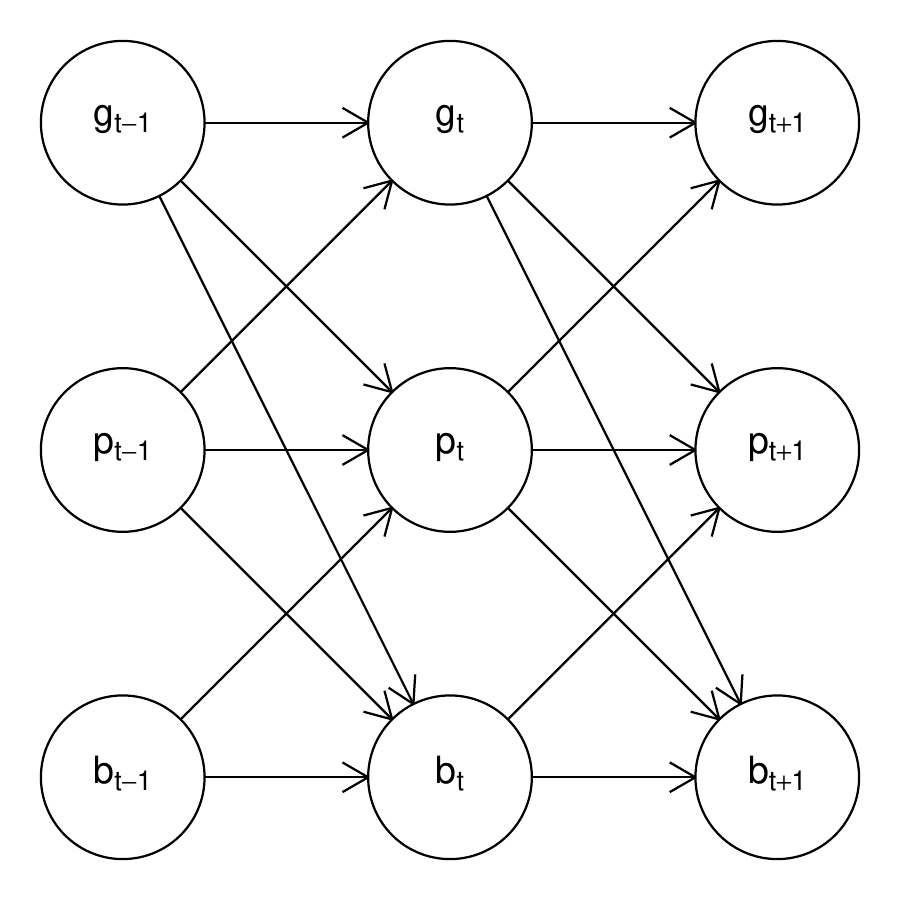} \caption{DAGs for the multivariate model created using the \fct{plot} method for \class{dynamitefit} objects. The left panel shows the model structure including the auxiliary response variable \code{logp} while the right panel shows the model structure where the auxiliary variable is not included. The latter DAG is obtained via a functional projection where the parents of \code{logp} become the parents of the children of \code{logp} and \code{logp} is removed from the graph at each time point.}\label{fig:multichanneldagplot}
\end{figure}
\end{CodeChunk}
Above, we used the argument \code{show_auxiliary} to project out the deterministic auxiliary variable \code{logp} from the DAG shown in the right panel of Figure~\ref{fig:multichanneldagplot}, which produces the same DAG as shown in Figure~\ref{fig:multichannel_dag}. In addition, the argument \code{show_covariates} can be used to control whether exogenous covariates should be included in the plot (the default is \code{FALSE} hiding covariates). Vertical, horizontal, and diagonal edges that would otherwise pass through vertices are automatically curved in the resulting figure to avoid overlapping with the vertices, but this can still occur with more complicated models.

To generate publication-quality figures with vector graphics, the argument \code{tikz} is provided. By setting \code{tikz = TRUE}, we can obtain the corresponding \pkg{TikZ} code for the figure as follows:
\begin{CodeChunk}
\begin{CodeInput}
R> cat(plot(multi_formula, show_auxiliary = FALSE, tikz = TRUE))
\end{CodeInput}
\begin{CodeOutput}
% Preamble
\usepackage{tikz}
\usetikzlibrary{positioning, arrows.meta, shapes.geometric}
\tikzset{%
  semithick,
  >={Stealth[width=1.5mm,length=2mm]},
  obs/.style 2 args = {
    name = #1, circle, draw, inner sep = 8pt, label = center:$#2$
  }
}
% DAG
\begin{tikzpicture}
  \node [obs = {v1}{g_{t - 1}}] at (-1, 3) {\vphantom{0}};
  \node [obs = {v2}{p_{t - 1}}] at (-1, 2) {\vphantom{0}};
  \node [obs = {v3}{b_{t - 1}}] at (-1, 1) {\vphantom{0}};
  \node [obs = {v4}{g_{t + 1}}] at (1, 3) {\vphantom{0}};
  \node [obs = {v5}{p_{t + 1}}] at (1, 2) {\vphantom{0}};
  \node [obs = {v6}{b_{t + 1}}] at (1, 1) {\vphantom{0}};
  \node [obs = {v7}{g_{t}}] at (0, 3) {\vphantom{0}};
  \node [obs = {v8}{p_{t}}] at (0, 2) {\vphantom{0}};
  \node [obs = {v9}{b_{t}}] at (0, 1) {\vphantom{0}};
  \draw [->] (v1) -- (v7);
  \draw [->] (v1) -- (v8);
  \draw [->] (v3) -- (v8);
  \draw [->] (v3) -- (v9);
  \draw [->] (v1) -- (v9);
  \draw [->] (v2) -- (v7);
  \draw [->] (v2) -- (v8);
  \draw [->] (v2) -- (v9);
  \draw [->] (v7) -- (v4);
  \draw [->] (v7) -- (v5);
  \draw [->] (v9) -- (v5);
  \draw [->] (v9) -- (v6);
  \draw [->] (v7) -- (v6);
  \draw [->] (v8) -- (v4);
  \draw [->] (v8) -- (v5);
  \draw [->] (v8) -- (v6);
\end{tikzpicture}
\end{CodeOutput}
\end{CodeChunk}
The default style used in the generated \pkg{TikZ} code mimics the style used in Figure~\ref{fig:multichannel_dag}.

\section{Model fitting and posterior inference}\label{sec:fitting}

To estimate the model, the declared model formula is supplied to the \fct{dynamite} function, which has the following arguments:
\begin{Code}
dynamite(
  dformula, data, time, group = NULL, priors = NULL, backend = "rstan",
  verbose = TRUE, verbose_stan = FALSE, stanc_options = list("O0"),
  threads_per_chain = 1L, grainsize = NULL, custom_stan_model = NULL,
  debug = NULL, ...
)
\end{Code}

This function parses the model formula and the data to generate a custom \proglang{Stan} model, which is then compiled and used to simulate the posterior distribution of the model parameters. The first three arguments of the function are mandatory. The first argument \code{dformula} is a \class{dynamiteformula} object that defines the model using the model components described in Section~\ref{sec:construction}. The second argument \code{data} is a \class{data.frame} or a \class{data.table} object that contains the variables used in the model formula. The third argument \code{time} is a column name of \code{data} that specifies the unique time points.

The remaining arguments of the function are optional. The \code{group} argument is a column name of \code{data} that specifies the unique groups (individuals), and when \code{group} is \code{NULL} we assume that there is only a single group (or individual). The argument \code{priors} supplies user-defined priors for the model parameters. The \proglang{Stan} backend can be selected using the \code{backend} argument, which accepts either \code{"rstan"} (the default) or \code{"cmdstanr"}. These options correspond to using the \pkg{rstan} and \pkg{cmdstanr} packages for the estimation, respectively. The \code{verbose} and \code{verbose_stan} arguments control the verbosity of the output from \fct{dynamite} and \proglang{Stan}, respectively. Additional \proglang{C++} compiler options such as the optimization level can be specified with \code{stanc_options} when using the \code{"cmdstanr"} backend.

While \proglang{Stan} supports between-chain parallelization via the \code{cores} and \code{parallel_chains} arguments for the \code{"rstan"} and \code{"cmdstanr"} backends, respectively, it also supports within-chain parallelization. In between-chain parallelization, the computations are split such that a single process is assigned one or more Markov-chains whereas in within-chain parallelization, the computations related to a single Markov chain are split, such as conditionally independent likelihood function evaluations. Both forms of parallelization can be leveraged via \fct{dynamite}. For between-chain parallelization, the \code{cores} and \code{parallel_chains} arguments can be passed directly to the backend sampling function via \code{...} (either \fct{rstan::sampling} or the \fct{sample} method of the \class{CmdStanModel} model object). For within-chain parallelization, threaded variants of all likelihood functions have been implemented in \pkg{dynamite} for the reduce-sum functionality of \proglang{Stan}, and the following two arguments are provided: \code{threads_per_chain} controls the number of threads to use per chain, and \code{grainsize} defines the suggested size of the partial sums (see the \proglang{Stan} manual for further information).

A custom \proglang{Stan} model code can be provided via \code{custom_stan_model}, which can be either a \class{character} string containing the model code or a path to a \code{.stan} file that contains the model code. Using this argument will override the automatically generated model code and it is intended for expert users only. Model customization is discussed at greater length in the related package vignette that can be accessed by writing \code{vignette("dynamite_custom", package = "dynamite")}. The \code{debug} argument can be used for various debugging options (see \code{?dynamite} for further information on these options and other arguments of the function).

The \code{data} argument should be supplied in long format, i.e., with \(N \times T\) rows in case of balanced panel data. Acceptable column types of \code{data} are \class{integer}, \class{logical}, \class{double}, \class{character}, objects of class \class{factor}, and objects of class \class{ordered factor}. Columns of the \class{character} type will be converted to \class{factor} columns. Beyond these standard types, any special classes such as \class{Date} whose internal storage type is one of the aforementioned types can be used, but these classes will be dropped, and the columns will be converted to their respective storage types. List columns are not supported. The \code{time} argument should be a \class{numeric} or a \class{factor} column of \code{data}. If \code{time} is a \class{factor} column, it will be converted to an \class{integer} column. Missing values in both response and predictor columns are supported but non-finite values are not. Observations with missing covariate or response values are omitted from the data when the model is fitted.

As an example, the following function call would estimate the model using data in the data frame \code{d}, which contains the variables \code{year} and \code{id} (defining the time-index and group-index variables of the data, respectively). Arguments \code{chains} and \code{cores} are passed to \fct{rstan::sampling} which then uses two parallel Markov chains in the MCMC sampling of the model parameters (as defined by \code{chains = 2} and \code{cores = 2}).
\begin{Code}
dynamite(
  dformula = obs(x ~ varying(~ -1 + w), family = "poisson") +
    splines(df = 10),
  data = d, time = "year", group = "id",
  chains = 2, cores = 2
)
\end{Code}
The output of \fct{dynamite} is a \class{dynamitefit} object for which the standard \proglang{S3} methods such as \fct{summary}, \fct{plot}, \fct{print}, \fct{fitted}, and \fct{predict} are provided along with various other methods and utility functions which we will describe in the following sections in more detail.

\subsection{User-defined priors}

The function \fct{get_priors} can be used to determine the parameters of the model whose prior distribution can be customized. The function can be applied to an existing model fit object (\class{dynamitefit}) or a model formula object (\class{dynamiteformula}). The function returns a \class{data.frame} object, which the user can then manipulate to include their desired priors and subsequently supply to the model fitting function \fct{dynamite}. The rationale behind the default prior specifications is discussed in detail in the related package vignette which can be viewed by writing \code{vignette("dynamite_priors", package = "dynamite")}.

For instance, using the model fit object \code{gaussian_example_fit} available in the \pkg{dynamite} package, we have the following priors:
\begin{CodeChunk}
\begin{CodeInput}
R> get_priors(gaussian_example_fit)
\end{CodeInput}
\begin{CodeOutput}
         parameter response             prior      type category
1 sigma_nu_y_alpha        y    normal(0, 3.1)  sigma_nu         
2          alpha_y        y  normal(1.5, 3.1)     alpha         
3      tau_alpha_y        y    normal(0, 3.1) tau_alpha         
4         beta_y_z        y    normal(0, 3.1)      beta         
5        delta_y_x        y    normal(0, 3.1)     delta         
6   delta_y_y_lag1        y    normal(0, 1.8)     delta         
7          tau_y_x        y    normal(0, 3.1)       tau         
8     tau_y_y_lag1        y    normal(0, 1.8)       tau         
9          sigma_y        y exponential(0.65)     sigma         
\end{CodeOutput}
\end{CodeChunk}
To customize a prior distribution, the user only needs to manipulate the \code{prior} column of the desired parameters in this \class{data.frame} using the appropriate \proglang{Stan} syntax and parametrization. For a categorical response variable, the column \code{category} describes which category the parameter is related to. For model parameters of the same type and response, a vectorized form of the corresponding distribution is automatically used in the generated \proglang{Stan} code if applicable. The definitions of the prior distributions are checked for validity before the model fitting process.

\subsection{Extracting model fit information}

We can obtain a simple model summary with the \fct{print} method of objects of class \class{dynamitefit}. For instance, the model fit object \code{gaussian_example_fit} gives the following output:
\begin{CodeChunk}
\begin{CodeInput}
R> print(gaussian_example_fit)
\end{CodeInput}
\begin{CodeOutput}
Model:
  Family   Formula                                       
y gaussian y ~ -1 + z + varying(~x + lag(y)) + random(~1)

Correlated random effects added for response(s): y

Data: gaussian_example (Number of observations: 1450)
Grouping variable: id (Number of groups: 50)
Time index variable: time (Number of time points: 30)

NUTS sampler diagnostics:

No divergences, saturated max treedepths or low E-BFMIs.

Smallest bulk-ESS: 137 (tau_alpha_y)
Smallest tail-ESS: 91 (sigma_y)
Largest Rhat: 1.007 (tau_alpha_y)

Elapsed time (seconds):
        warmup sample
chain:1 10.255  5.763
chain:2 18.894 10.197

Summary statistics of the time- and group-invariant parameters:
# A tibble: 6 x 10
  variable     mean median      sd     mad     q5   q95  rhat ess_bulk
  <chr>       <dbl>  <dbl>   <dbl>   <dbl>  <dbl> <dbl> <dbl>    <dbl>
1 beta_y_z   1.97   1.97   0.0122  0.0120  1.95   1.99  0.998     214.
2 sigma_nu_~ 0.0946 0.0940 0.0106  0.0107  0.0791 0.113 1.00      199.
3 sigma_y    0.198  0.198  0.00397 0.00349 0.192  0.205 0.998     148.
4 tau_alpha~ 0.203  0.195  0.0482  0.0434  0.139  0.292 1.01      137.
5 tau_y_x    0.368  0.355  0.0746  0.0671  0.257  0.508 1.00      195.
6 tau_y_y_l~ 0.104  0.101  0.0202  0.0215  0.0767 0.139 1.00      196.
# i 1 more variable: ess_tail <dbl>
\end{CodeOutput}
\end{CodeChunk}
By default, the argument \code{full_diagnostics} of the \fct{print} method is set to \code{FALSE} which means that the model diagnostics are computed only for the time-invariant and non-group-specific parameters. Setting this argument to \code{TRUE} will compute the diagnostics for all model parameters which can be time-consuming for complex models. Convergence of the MCMC chains and the smallest effective sample sizes of the model parameters can be assessed using the \fct{mcmc_diagnostics} method of \class{dynamitefit} object whose arguments are the model fit object and \code{n}, the number of potentially problematic variables to report (default is 3). We refer the reader to \citep{vehtari2021rhat} and to the documentation of the \fct{rstan::check_hmc_diagnostics} and \fct{posterior::default_convergence_measures} functions for detailed information on the diagnostics reported by the \fct{mcmc_diagnostics} function.
\begin{CodeChunk}
\begin{CodeInput}
R> mcmc_diagnostics(gaussian_example_fit)
\end{CodeInput}
\begin{CodeOutput}
NUTS sampler diagnostics:

No divergences, saturated max treedepths or low E-BFMIs.

Smallest bulk-ESS values: 
                
alpha_y[28]   72
alpha_y[10]  126
delta_y_x[7] 126

Smallest tail-ESS values: 
                 
nu_y_alpha_id6 83
sigma_y        91
alpha_y[28]    94

Largest Rhat values: 
                       
delta_y_y_lag1[28] 1.03
alpha_y[29]        1.03
alpha_y[28]        1.03
\end{CodeOutput}
\end{CodeChunk}
We note that due to CRAN file size restrictions, the number of stored posterior samples in this example \class{dynamitefit} object is very small, leading to small effective sample sizes. Diagnostics specific to HMC can be extracted with the \fct{hmc_diagnostics} method.

A table of posterior draws or summaries of each parameter of the model can be obtained with the methods \fct{as.data.frame} and \fct{as.data.table} which differ only by their output type (\class{data.frame} and \class{data.table}). More specifically, the output of \fct{as.data.frame} is a tibble; a tidyverse variant of data frames of class \class{tbl_df} as defined in the \pkg{tibble} package \citep{tibble}. These two methods have the following arguments:
\begin{Code}
as.data.frame.dynamitefit(
  x, keep.rownames, row.names = NULL, optional = FALSE, types = NULL,
  parameters = NULL, responses = NULL, times = NULL, groups = NULL,
  summary = FALSE, probs = c(0.05, 0.95), include_fixed = TRUE, ...
)
\end{Code}
Here, \code{x} is the \class{dynamitefit} object and \code{types} is a \class{character} vector that determines the types parameters that will be included in the output. If \code{types} is not used, a \class{character} vector argument \code{parameters} can be used to specify exactly which parameters of the model should be included. The argument \code{responses} can be used select parameters that are related to specific response variables. For determining suitable options for the arguments \code{types} and \code{parameters}, methods \fct{get_parameter_types} and \fct{get_parameter_names} can be used. The arguments \code{times} and \code{groups} can be used to further restrict the parameters in the output to only include specific time points or groups, respectively. The argument \code{summary} determines whether to provide summary statistics (mean, standard deviation, and quantiles selected by the argument \code{probs}) of each parameter, or the full posterior draws. The argument \code{include_fixed} determines whether to include parameters related to fixed time points in the output (see Section~\ref{sec:lags} for details on fixed time points). The default arguments of the methods \code{keep.rownames}, \code{row.names}, \code{optional}, and \code{...} are ignored for \class{dynamitefit} objects. All parameter types used in \pkg{dynamite} are described in Table~\ref{tab:dynamite_types}.

\begin{table}[!ht]
  \centering
  \begin{tabular}{ll}
    \toprule
    Parameter type        & Description \\
    \midrule
    \code{"alpha"}        & Intercept terms (time-invariant \(\alpha_{c}\) or time-varying \(\alpha_{c,t}\)) \\
    \code{"beta"}         & Time-invariant regression coefficients \(\beta_c\) \\
    \code{"corr"}         & Pairwise correlations of multivariate Gaussian responses \\
    \code{"corr_nu"}      & Pairwise within-group correlations of random effects \(\nu_{c,i}\) \\
    \code{"corr_psi"}     & Pairwise correlations of the latent factors \(\psi_{c,t}\) \\
    \code{"cutpoint"}     & Cutpoints for ordinal regression (time-invariant or time-varying) \\
    \code{"delta"}        & Time-varying regression coefficients \(\delta_{c,t}\) \\
    \code{"kappa"}        & The contribution of latent factor loadings in the total variation \\
    \code{"lambda"}       & Latent factor loadings \(\lambda_{c,i}\) of the latent factors \(\psi_{c,t}\) \\
    \code{"nu"}           & Group-level random effects \(\nu_{c,i}\) \\
    \code{"omega"}        & Spline coefficients \(\omega_{c,k}\) of the regression coefficients \(\delta_{c,t}\) \\
    \code{"omega_alpha"}  & Spline coefficients of the time-varying intercepts \(\alpha_{c,t}\) \\
    \code{"omega_psi"}    & Spline coefficients of the latent factors \(\psi_{c,t}\) \\
    \code{"phi"}          & Describes various distributional parameters, such as: \\
                          & the dispersion parameter of the negative binomial distribution, \\
                          & the shape parameter of the gamma distribution, \\
                          & the precision parameter of the beta distribution, \\
                          & the degrees of freedom of the Student \(t\) distribution. \\
    \code{"psi"}          & Latent factors \(\psi_{c,t}\) \\
    \code{"sigma"}        & Standard deviations of (multivariate) Gaussian responses \\
    \code{"sigma_lambda"} & Standard deviations of the latent factor loadings \(\lambda_{c,i}\) \\
    \code{"sigma_nu"}     & Standard deviations of the random effects \(\nu_{c,i}\) \\
    \code{"tau"}          & Standard deviations \(\tau_{c,k}\) of \(\omega_{c,k,d}\) \\
    \code{"tau_alpha"}    & Standard deviations of the spline coefficients of \(\alpha_{c,t}\) \\
    \code{"tau_psi"}      & Standard deviations of the spline coefficients of \(\psi_{c,t}\) \\
    \code{"zeta"}         & Total variation of latent factors, i.e., \(\sigma_\lambda + \tau_\psi\) \\
    \bottomrule
  \end{tabular}
  \caption{The parameter types used in \pkg{dynamite}.}
  \label{tab:dynamite_types}
\end{table}

For instance, we can extract the posterior summary of the time-invariant regression coefficients (\code{types = "beta"}) for the response variable \code{y} in the \code{gaussian_example_fit} object by writing:
\begin{CodeChunk}
\begin{CodeInput}
R> as.data.frame(
+    gaussian_example_fit,
+    responses = "y", types = "beta", summary = TRUE
+  )
\end{CodeInput}
\begin{CodeOutput}
# A tibble: 1 x 10
  parameter  mean     sd    q5   q95  time group category response
  <chr>     <dbl>  <dbl> <dbl> <dbl> <int> <int> <chr>    <chr>   
1 beta_y_z   1.97 0.0122  1.95  1.99    NA    NA <NA>     y       
# i 1 more variable: type <chr>
\end{CodeOutput}
\end{CodeChunk}
For \class{dynamitefit} objects, the \fct{summary} method is a shortcut for \code{as.data.frame(summary = TRUE)}.

The generated \proglang{Stan} code of the model can be extracted with the method \fct{get_code} as a \class{character} string. This feature is geared towards advanced users who may for example need to make slight modifications to the generated code in order to adapt the model to a specific scenario that cannot be accomplished with the \pkg{dynamite} model syntax. The generated code also contains helpful annotations describing the model blocks, parameters, and complicated code sections. Using the argument \code{blocks}, we can extract only specific blocks of the full model code. To illustrate, we extract the parameters block of the \code{gaussian_example_fit} model code as the full model code is too large to display.
\begin{CodeChunk}
\begin{CodeInput}
R> cat(get_code(gaussian_example_fit, blocks = "parameters"))
\end{CodeInput}
\begin{CodeOutput}
parameters {
  // Random group-level effects
  vector<lower=0>[M] sigma_nu; // standard deviations of random effects
  matrix[N, M] nu_raw;
  vector[K_fixed_y] beta_y; // Fixed coefficients
  matrix[K_varying_y, D] omega_y; // Spline coefficients
  vector<lower=0>[K_varying_y] tau_y; // SDs for the random walks
  real a_y; // Mean of the first time point
  row_vector[D - 1] omega_raw_alpha_y; // Coefficients for alpha
  real<lower=0> tau_alpha_y; // SD for the random walk
  real<lower=0> sigma_y; // SD of the normal distribution
}
\end{CodeOutput}
\end{CodeChunk}
Conversely, a customized \proglang{Stan} model code can be supplied to \fct{dynamite} using the \code{custom_stan_model} argument.

\subsection{Visualizing the posterior distributions}

The \fct{plot} method for \class{dynamitefit} objects can be used to obtain plots of various types of the model fit using the \pkg{ggplot2} package to produce the plots. This method has the following arguments:
\begin{Code}
plot.dynamitefit(
  x, plot_type = c("default", "trace", "dag"), types = NULL,
  parameters = NULL, responses = NULL, groups = NULL, times = NULL,
  level = 0.05, alpha = 0.5, facet = TRUE, scales = c("fixed", "free"),
  n_params = NULL, ...
)
\end{Code}
The arguments \code{type}, \code{parameters}, \code{responses}, \code{groups} and \code{times} are analogous to those of the \fct{as.data.frame} method for selecting which parameters should be plotted.
Arguments \code{level}, \code{alpha}, \code{facet} and \code{scales} control the visual aspects of the plot: \code{level} defines the plotted posterior intervals as \code{100 * (1 - 2 * level)} \% intervals, \code{alpha} is the opacity level for \fct{ggplot2::geom_ribbon} for plotting posterior intervals, \code{facet} determines whether time-invariant parameters should be plotted together (\code{FALSE}) or separately using \fct{ggplot2::facet_wrap} (\code{TRUE}), and \code{scales} selects whether the vertical axis of different parameters should be the same (\code{"fixed"}) or allowed to vary between parameters (\code{"free"}). Finally, \code{n_params} controls the maximum number of parameters of each type to plot. By default, the number of parameters is limited to prevent accidental plots with a large number of parameters that may take an excessively long time to render. Next, we showcase some example plots and the different plot types that are available via the \code{plot_type} argument.

For instance, Figure~\ref{fig:parameter_posterior_plot} shows the posterior means and posterior intervals of the time-varying intercept (type \code{"alpha"}) and time-varying regression coefficients (type \code{"delta"}) in the \code{gaussian_example_fit} model (using the \code{"default"} option of the \code{plot_type} argument by default).
\begin{CodeChunk}
\begin{CodeInput}
R> plot(
+    gaussian_example_fit,
+    types = c("alpha", "delta"), scales = "free"
+  ) +
+    labs(title = "")
\end{CodeInput}
\begin{figure}
\includegraphics[width=\maxwidth]{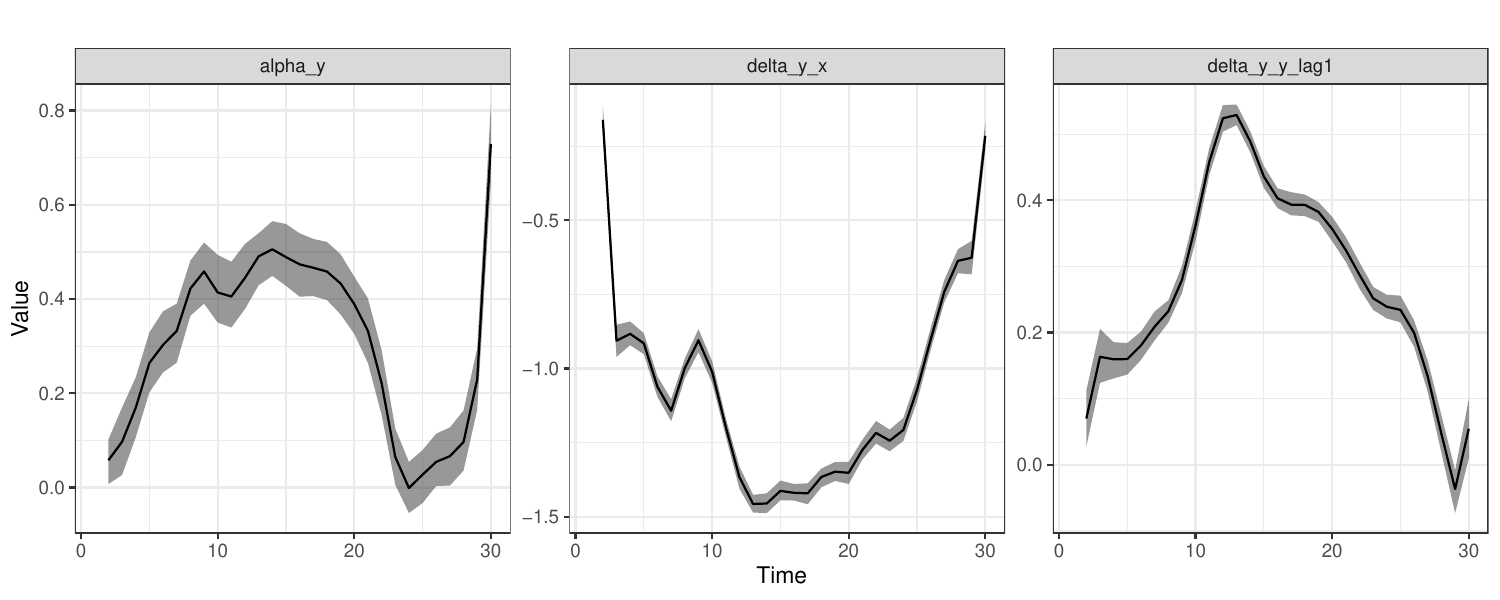} \caption{Posterior means (black lines) and 90\% posterior intervals (gray areas) for the time-varying coefficients for the response variable \code{y} in the \code{gaussian\_example\_fit} model. The panels from left to right show the time-varying intercept for \code{y}, the time-varying effect of \code{x} on \code{y}, and the time-varying effect of \code{lag(y)} (the previous time-point) on \code{y}.}\label{fig:parameter_posterior_plot}
\end{figure}
\end{CodeChunk}
While \code{plot_type = "default"} produces plots such as Figure~\ref{fig:parameter_posterior_plot}, using \code{plot_type = "trace"} instead provides the marginal posterior densities and traceplots of the MCMC chains, as shown in Figure~\ref{fig:gaussian_trace} where we also select the time-invariant regression coefficients of the model to be plotted.
\begin{CodeChunk}
\begin{CodeInput}
R> plot(gaussian_example_fit, plot_type = "trace", types = "beta")
\end{CodeInput}
\begin{figure}
\includegraphics[width=\maxwidth]{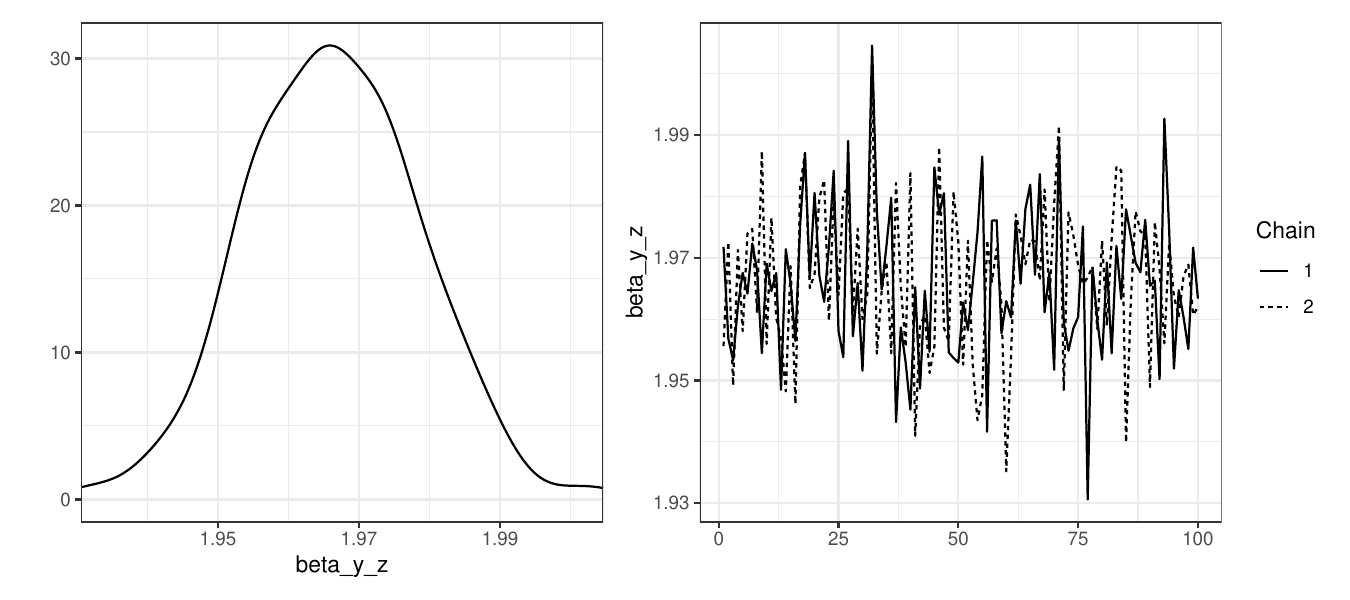} \caption[Marginal posterior density and traceplot of the MCMC chains of the time-invariant regression coefficient \code{beta\_y\_z} of \code{z} for the response variable \code{y} in the \code{gaussian\_example\_fit} model]{Marginal posterior density and traceplot of the MCMC chains of the time-invariant regression coefficient \code{beta\_y\_z} of \code{z} for the response variable \code{y} in the \code{gaussian\_example\_fit} model.}\label{fig:gaussian_trace}
\end{figure}
\end{CodeChunk}
The third option \code{plot_type = "dag"} can be used to visualize the structure of the model as a DAG as shown in Figure~\ref{fig:multichanneldagplot} and described in Section~\ref{sec:modelvis}.

\subsection{Missing data and multiple imputation}

Panel data often contains missing observations for various reasons. A common approach in a Bayesian setting is to treat missing observations as additional unknown parameters, and to sample them along with the model parameters during MCMC. However, the MCMC sampling in \pkg{dynamite} is based on \proglang{Stan}'s variant of the gradient-based NUTS algorithm \citep{hoffman2014, betancourt2018}, which cannot be used to sample discrete variables such as missing count data. Therefore, the default behavior in \pkg{dynamite} is to use a complete-case approach which is unbiased when data are missing completely at random as well as in certain other specific settings \citep{vanBuuren2018}. As an alternative to complete-case analysis with \fct{dynamite}, the function \fct{dynamice} first performs multiple imputation using the imputation algorithms of the \pkg{mice} package \citep{vanBuuren2011}, runs MCMC on each imputed sample, and combines the posterior samples of each run, as suggested for example in \citep{bdabook}.

The \fct{dynamice} function has all of the arguments of \fct{dynamite} with some additions. The argument \code{mice_args} is a \class{list} that can be used to provide arguments to the underlying imputation function \fct{mice} of the \pkg{mice} package. Format of the data during imputation can be selected with the \code{impute_format} argument that accepts either \code{"wide"} or \code{"long"}. Data in wide format will have one group per row (with observations at different time points in different columns) while data in long format corresponds to the standard data format of \fct{dynamite} described in Section~\ref{sec:fitting}. Argument \code{keep_imputed} is a \class{logical} value can be used to select whether the imputed data sets should be included in the return object of \fct{dynamice}. If \code{TRUE}, the imputed data sets will be found in the \code{imputed} field of the returned \class{dynamitefit} object. All of the methods for \class{dynamitefit} objects are available also for model fits obtained from \fct{dynamice}, but it should be noted that convergence measures and effective samples sizes such as those reported by \fct{mcmc_diagnostics} may be unreliable for such model fits.

\section{Prediction}\label{sec:prediction}

The \pkg{dynamite} package provides a comprehensive set of features for obtaining predictions based on the posterior distribution of the model parameters. The package supports the imputation of missing exogenous covariate values (via last observation carried forward or next observation carried backward), aggregated and individual-level predictions, and various methods to account for new levels of the \code{group} variable for random effects. Counterfactual predictions can also be obtained which enables the study of causal effects and other intricate causal quantities. It should be noted that the predictions do not directly support forecasting as there is no unambiguous way to define how the splines for the time-varying regression coefficients should behave outside of the observed time points. However, such predictions can be obtained by augmenting the original data with missing values for future time points. Furthermore, the package can be used to generate data from a DMPM without an existing model fit by first specifying the values of the model parameters and the fixed covariates (see the package vignette on data simulation for further information: \code{vignette("dynamite_simulation", package = "dynamite")}).

The \fct{predict} method for \class{dynamitefit} objects can be used to obtain predictions from the posterior predictive distribution. This function has the following arguments:
\begin{Code}
predict.dynamitefit(
  object, newdata = NULL, type = c("response", "mean", "link"),
  funs = list(), impute = c("none", "locf", "nocb"),
  new_levels = c("none", "bootstrap", "gaussian", "original"),
  global_fixed = FALSE, n_draws = NULL, thin = 1,
  expand = TRUE, df = TRUE, ...
)
\end{Code}
We will only explain the most important arguments of this method and refer the reader to the package documentation for more information. The first argument \code{object} is the \class{dynamitefit} object that the predictions will be based on. The argument \code{newdata} can be used to define the groups, time points, and covariate values that the predictions should be computed for. If \code{newdata} is \code{NULL}, predictions will be computed for the original \code{data} supplied to the \fct{dynamite} function when the model was fitted from the first non-fixed time point onward. The \code{type} argument selects the type of computed predictions. By default, \code{type = "response"} returns the individual-level simulated predictions for the response variables of the model. Options \code{"link"} and \code{"mean"} return the linear predictor values and the expected values of the posterior predictive distribution, respectively. The argument \code{n_draws} controls the number of posterior draws to be used for prediction. By default, all draws are used. Alternatively, the argument \code{thin} can be used to select every \code{thin}th posterior draw to be used for the prediction task.

For example, we can obtain posterior predictive samples for the first 4 groups in the \code{gaussian_example} dataset using the corresponding model fit object \code{gaussian_example_fit} with the first 50 posterior draws. The predictions are shown in Figure~\ref{fig:gaussianpred} and can be obtained as follows:
\begin{CodeChunk}
\begin{CodeInput}
R> pred <- predict(gaussian_example_fit, n_draws = 50)
R> pred |>
+    dplyr::filter(id < 5) |>
+    ggplot(aes(time, y_new, group = .draw)) +
+    geom_line(alpha = 0.5) +
+    geom_line(aes(y = y), colour = "tomato") +
+    facet_wrap(~ id)
\end{CodeInput}
\begin{figure}
\includegraphics[width=\maxwidth]{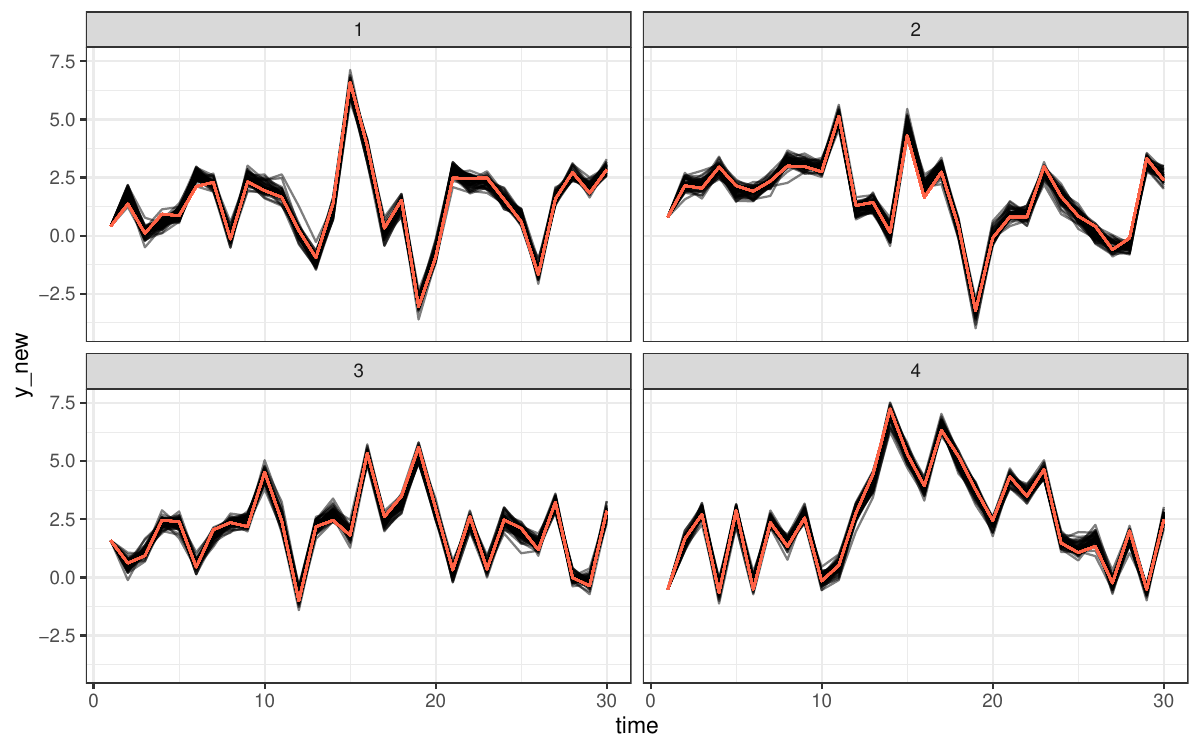} \caption[Posterior predictive samples for the first 4 groups of the \code{gaussian\_example} data]{Posterior predictive samples for the first 4 groups of the \code{gaussian\_example} data. Lines in red represent the observed values.}\label{fig:gaussianpred}
\end{figure}
\end{CodeChunk}
The \fct{fitted} method is also provided for \class{dynamitefit} objects. In contrast to multi-step predictions of \fct{predict}, this function computes expected values of the posterior predictive distributions at each time point conditional on the original observations.

We note that the multi-step predictions contain not only the parameter uncertainty but also the inherent aleatoric (stochastic) uncertainty of the trajectories. The Monte Carlo variation due to the finite number of posterior samples can be reduced by increasing the number of iterations or chains of the MCMC run (as with any posterior summaries) or by combining samples from multiple \fct{predict} calls in case the Monte Carlo error is mostly due to the trajectory simulation.

\subsection{Aggregated predictions and memory conservation}\label{sec:summary_prediction}

For large datasets and complicated models, obtaining individual-level predictions can be memory-intensive. For example, data with 100 groups, 100 time points, a categorical response with 4 categories, and 1000 posterior draws would result in 40 million elements. A simple way to reduce memory usage is to set the argument \code{expand} of \fct{predict} to \code{FALSE} (the default is \code{TRUE}). Disabling this argument separates the simulated values from the fixed covariates in the model into two \class{data.table} objects in the output, called \code{simulated} and \code{observed}, which are then returned as a \class{list} object. This optimization is always carried out internally, meaning that the value of the \code{expand} argument only affects the returned output.

To further reduce memory usage, the argument \code{funs} can be used to obtain aggregated predictions instead of the full individual-level predictions for each time point. This argument accepts a named list of lists of named functions for each response variable of the model, where the supplied functions are then applied over the individuals. The resulting columns in the output are named based on the function names and the response variables. The \code{expand} argument is automatically set to \code{FALSE} when using the \code{funs} argument. For example, we could compute the mean and standard deviation of the predictions for the response variable \code{y} in the \code{gaussian_example} dataset at each time point as follows:
\begin{CodeChunk}
\begin{CodeInput}
R> pred_funs <- predict(
+    gaussian_example_fit,
+    funs = list(y = list(mean = mean, sd = sd))
+  )
R> head(pred_funs$simulated)
\end{CodeInput}
\begin{CodeOutput}
    mean_y      sd_y time .draw
1       NA        NA    1     1
2 1.515636 0.8815666    2     1
3 1.667627 1.3117554    3     1
4 1.743166 1.2392450    4     1
5 2.151360 1.2395050    5     1
6 2.218212 1.3884278    6     1
\end{CodeOutput}
\end{CodeChunk}
The reduction in memory usage compared to the full individual-level predictions is rather substantial even in this simple scenario:
\begin{CodeChunk}
\begin{CodeInput}
R> library("pryr")
R> pred_full <- predict(gaussian_example_fit)
R> object_size(pred_full)
\end{CodeInput}
\begin{CodeOutput}
12.00 MB
\end{CodeOutput}
\begin{CodeInput}
R> object_size(pred_funs)
\end{CodeInput}
\begin{CodeOutput}
188.34 kB
\end{CodeOutput}
\end{CodeChunk}
The \code{funs} argument can also be used to aggregate the expected values of the posterior predictive distribution with \code{type = "mean"}:
\begin{CodeChunk}
\begin{CodeInput}
R> pred_funs_mean <- predict(
+    gaussian_example_fit,
+    type = "mean",
+    funs = list(y = list(mean = mean, sd = sd))
+  )
R> head(pred_funs_mean$simulated)
\end{CodeInput}
\begin{CodeOutput}
    mean_y      sd_y time .draw
1       NA        NA    1     1
2 1.498594 0.8395451    2     1
3 1.667819 1.3096773    3     1
4 1.746313 1.2035324    4     1
5 2.138912 1.2168148    5     1
6 2.208098 1.3451194    6     1
\end{CodeOutput}
\end{CodeChunk}

\section{Summary} \label{sec:summary}

In this paper, we presented the \pkg{dynamite} package for Bayesian inference of DMPMs. The package provides a user-friendly interface for model construction, estimation, prediction, posterior inference, and visualization with extensive and detailed documentation of its features. The package has been designed to be as general as possible by supporting multivariate models, many response variable distributions, custom prior distributions, and common model features such as time-varying effects and group-specific random effects. The package design also aims for high performance in model estimation by employing \proglang{Stan} and in general-purpose data manipulation by using \pkg{data.table} which is especially reflected in prediction. For advanced users, the \proglang{Stan} code generated by \pkg{dynamite} can be extracted and adapted to user-specific scenarios.

In the future, we plan to extend the capabilities of \pkg{dynamite} by adding support for more distributions. Some distributions in \proglang{Stan} also lack efficient likelihood function variants, such as the Bernoulli distribution, which will likely become available in the future and will be subsequently implemented in \pkg{dynamite} as well.

\section*{Acknowledgments}

This research was funded by the Research Council of Finland (decision numbers 331817, 355153, 345546) and partially supported by the INVEST Research Flagship Centre.

\bibliography{dynamite}

\appendix
\section{Details on latent factors}\label{app:latent_factor}

Latent factor models with product terms \(\lambda_i\psi_t\) are known to suffer from identifiability issues. For example, it is possible to multiply each \(\lambda_i\) by some constant \(c\) while simultaneously multiplying \(\psi_i, t=1,\ldots, T\) with the reciprocal of the same constant, leading to the same likelihood value as the original model. In case of multiple latent factors and (vector) autoregressive process on \(\psi_t\), \citet{bai2015} discuss two alternative identifiability constraints, which in our single factor model translate to fixing \(\lambda_i = 1\) for some \(i\), or constraining \(\lambda_i > 0\) for some \(i\), with an additional constraint that the standard deviation of the noise term of \(\psi_t\) is 1. In both cases, we need to decide which individual is used as a reference for the constrained \(\lambda_i\). This choice can lead to computational issues if the true value of \(\lambda_i\) is not compatible with these constrains (e.g., the true value is zero). Instead, we define the constraints via the mean of \(\lambda\).

Denote the expected value of the factor loadings as \(\bar \lambda\). Now write \(\lambda_i = \bar \lambda + \sigma^\ast_\lambda \lambda^\ast_i\) where \(\lambda^\ast_i \sim N(0, 1)\). While \pkg{dynamite} models \(\psi_t\) as spline, for the ease of exposition here we assume \(\psi_t\) is a simple random walk \(\psi_t = \psi_{t-1} + \sigma_\psi\xi_t\).

Assume first that \(\bar \lambda \neq 0\). In this case, we can write
\[
  (\bar\lambda + \sigma^\ast_\lambda \lambda^\ast_i)\psi_t, \quad \psi_t = \psi_{t-1} + \sigma_\psi\xi_t
\]
as
\[
  \lambda_i\psi_t, \quad \psi_t = \psi_{t-1} + \tau_\psi\xi_t,
\]
where \(\lambda_i = 1 + \sigma_\lambda\lambda_i^\ast\), \(\sigma_\lambda= \sigma^\ast_\lambda / \bar\lambda\), and \(\tau_\psi = \bar\lambda\sigma_\psi\). Sampling \(\sigma_\lambda\) and \(\tau_\psi\) can be inefficient due to the strong negative correlation between these parameters, so instead we sample (and set priors for) \(\zeta = \sigma_\lambda + \tau_\psi\) and \(0 < \kappa < 1\) so that \(\sigma_\lambda = \kappa \zeta\) and \(\tau_\psi = (1 - \kappa) \zeta\).

If instead \(\bar\lambda = 0\), then \(\lambda_i\psi_t = \sigma^\ast_\lambda \lambda^\ast_i\psi_t\) is not uniquely identifiable, so we fix \(\tau_\psi=1\) and sample \(\sigma_\lambda\) directly. However, it is still possible to encounter multimodality due to sign-switching, which does not affect the predictions obtained from the model, but the automatic diagnostics of MCMC samples can be misleading. By default, \pkg{dynamite} tries to fix this by adjusting the signs of the \(\lambda\) and \(\psi\) terms based on the mean of the spline coefficients corresponding to \(\psi\). However, this only works if the mean of the spline coefficients is not close to zero, and it is possible to turn this option off so that the user can try to fix the sign-switching in the post-processing steps, e.g., by using the algorithms of the \pkg{label.switching} package \citep{Papastamoulis2016}.

\end{document}